\documentclass[useAMS]{mn2e}

\usepackage{epsfig,natbib_vw,bm,times}

\bibpunct[, ]{(}{)}{;}{a}{}{,}

\def\bj{B_{\scriptscriptstyle\rm J}}
\def\rf{R_{\scriptscriptstyle\rm F}}

\def\be{\begin{equation}}
\def\ee{\end{equation}}

\def\gsim{\mathrel{\lower0.6ex\hbox{$\buildrel {\textstyle >}
 \over {\scriptstyle \sim}$}}}
\def\lsim{\mathrel{\lower0.6ex\hbox{$\buildrel {\textstyle <}
 \over {\scriptstyle \sim}$}}}

\def\m@th{\mathsurround=0pt }
\def\eqalign#1{\null\,\vcenter{\openup1\jot \m@th
 \ialign{\strut\hfil$\displaystyle{##}$&$\displaystyle{{}##}$\hfil
 \crcr#1\crcr}}\,}

\def \aj {AJ}
\def \mnras {MNRAS}
\def \apj {ApJ}
\def \apjs {ApJS}
\def \apjl {ApJL}

\def \nat {Nature}

\def \Mpc {{\, \rm Mpc}}

\def \deg {^{\circ}}
\def \mpcoh {\,h^{-1}\,{\rm Mpc}}
\def \japsub{\rm\scriptstyle}
\def \Ej{{\rm\scriptstyle E}}
\def \Lj{{\rm\scriptstyle L}}
\def \LN{{\rm LN}}

\title[Stochastic relative bias in the 2dFGRS]
      {The 2dF Galaxy Redshift Survey: stochastic
      relative biasing between galaxy populations}
\author[V. Wild et al.]{
\parbox[t]{\textwidth}{\raggedright
Vivienne Wild$^1$\thanks{vw@ast.cam.ac.uk},
John A.\ Peacock$^{2}$, 
Ofer Lahav$^{1, 16}$,
Edward Conway$^{11}$,
Steve Maddox$^{11}$,
Ivan K.\ Baldry$^9$,
Carlton M.\ Baugh$^6$,
Joss Bland-Hawthorn$^3$,
Terry Bridges$^3$, 
Russell Cannon$^3$, 
Shaun Cole$^6$, 
Matthew Colless$^3$, 
Chris Collins$^4$, 
Warrick Couch$^5$, 
Gavin Dalton$^{7,15}$,
Roberto De Propris$^{3,17}$,
Simon P.\ Driver$^{14}$, 
George Efstathiou$^1$, 
Richard S.\ Ellis$^8$, 
Carlos S.\ Frenk$^6$, 
Karl Glazebrook$^9$, 
Carole Jackson$^{14}$,
Ian Lewis$^7$, 
Stuart Lumsden$^{10}$, 
Darren Madgwick$^{12}$,
Peder Norberg$^{13}$,
Bruce A.\ Peterson$^{14}$, 
Will Sutherland$^{1}$,
Keith Taylor$^8$ (The 2dFGRS Team)}
\vspace*{6pt} \\ 
$^1$Institute of Astronomy, University of Cambridge, Madingley Road,
    Cambridge CB3 0HA, UK \\
$^2$Institute for Astronomy, University of Edinburgh, Royal Observatory, 
       Blackford Hill, Edinburgh EH9 3HJ, UK \\
$^3$Anglo-Australian Observatory, P.O.\ Box 296, Epping, NSW 2111,
    Australia\\  
$^4$Astrophysics Research Institute, Liverpool John Moores University,  
    Twelve Quays House, Birkenhead, L14 1LD, UK \\
$^5$Department of Astrophysics, University of New South Wales, Sydney, 
    NSW 2052, Australia \\
$^6$Department of Physics, University of Durham, South Road, 
    Durham DH1 3LE, UK \\ 
$^7$Department of Physics, University of Oxford, Keble Road, 
    Oxford OX1 3RH, UK \\
$^8$Department of Astronomy, California Institute of Technology, 
    Pasadena, CA 91025, USA \\
$^9$Department of Physics \& Astronomy, Johns Hopkins University,
       Baltimore, MD 21118-2686, USA \\
$^{10}$Department of Physics, University of Leeds, Woodhouse Lane,
       Leeds, LS2 9JT, UK \\
$^{11}$School of Physics \& Astronomy, University of Nottingham,
       Nottingham NG7 2RD, UK \\
$^{12}$Lawrence Berkeley National Laboratory, 1 Cyclotron Road,
       Berkeley, CA 94720, USA \\ 
$^{13}$ETHZ Institut f\"ur Astronomie, HPF G3.1, ETH H\"onggerberg, CH-8093
       Z\"urich, Switzerland \\
$^{14}$Research School of Astronomy \& Astrophysics, The Australian 
    National University, Weston Creek, ACT 2611, Australia \\
$^{15}$Rutherford Appleton Laboratory, Chilton, Didcot, OX11 0QX, UK
    \\
$^{16}$Department of Physics and Astronomy, University College London,
London, WC1E 6BT, UK \\
$^{17}$Astrophysics Group, Department of Physics, Bristol University,
Tyndall Avenue, Bristol, BS8 1TL, UK\\
$^{\star}$email: vw@ast.cam.ac.uk\\
}

\begin{document}
\maketitle
\begin{abstract}
It is well known that the clustering of galaxies depends on galaxy type.
Such {\it relative bias\/} complicates the inference of cosmological parameters 
from galaxy redshift surveys, and is a challenge to theories of galaxy formation
and evolution. In this paper we perform a joint counts-in-cells
analysis on galaxies in the 2dF Galaxy Redshift Survey, classified
by both colour and spectral type, $\eta$, as early or late type galaxies. 
We fit three different models of relative bias to the joint probability 
distribution of the cell counts, assuming Poisson sampling of 
the galaxy density field. We investigate the nonlinearity and stochasticity 
of the relative bias, with cubical cells of side $10\Mpc \leq L \leq
45\Mpc$ ($h=0.7$). 
Exact linear bias is ruled out with high significance on all scales. 
Power law bias gives a better fit, but likelihood ratios prefer a
bivariate lognormal distribution, with a 
non-zero `stochasticity' -- i.e. scatter that may result 
from physical effects on galaxy formation other than those from the local density field.
Using this model, we measure a correlation coefficient in log-density space ($r_{\LN}$)
of 0.958 for cells of length $L=10\Mpc$, increasing to 0.970 by $L=45\Mpc$.
This corresponds to a stochasticity $\smash{\sigma_b/\hat{b}}$ of
$0.44\pm0.02$ and $0.27\pm0.05$ respectively. For smaller cells, the
Poisson sampled lognormal distribution presents an increasingly poor fit
to the data, especially with regard to the fraction of completely empty
cells. We compare these trends with the predictions of semianalytic galaxy
formation models: these match the data well in terms of overall level of
stochasticity, variation with scale, and fraction of empty cells.
\end{abstract}


\section{Introduction}
The question of whether galaxies trace the matter distribution of the
universe has many implications for cosmology and galaxy formation theories.
Since \citet{1931ApJ....74...43H} it has been known that galaxies of different type
cluster differently, and as such it cannot be possible for all
galaxies to trace the matter distribution exactly. This observation
has been reconfirmed many times, traditionally by comparisons of the
correlation functions of different subgroups. For example, early type
(or passive) galaxies are more strongly clustered than late type (or
actively  star-forming) galaxies
(e.g. \citealt{1976ApJ...208...13D,1980ApJ...236..351D};
\citealt*{1990ApJ...350..119L};
\citealt{1996MNRAS.283..709H,2002MNRAS.332..827N,2002ApJ...571..172Z,2003MNRAS.344..847M})  
and luminous galaxies cluster more strongly than faint galaxies 
(e.g. \citealt*{1998AJ....115..869W};
\citealt{2001MNRAS.328...64N,2002MNRAS.332..827N,2002ApJ...571..172Z,2004ApJ...608...16Z}). 

Any difference in the distribution of galaxies relative to mass has
become known as {\it galaxy bias\/}. This assumed a central importance in cosmology
via the attempts to rescue the $\Omega_m = 1$ universe
after observations of cluster mass-to-light ratios suggested values
closer to $\Omega_m = 0.2$. Such bias could
occur if the galaxy formation efficiency were increased in overdense
regions of space, the so called `high peak scenario' 
\citep{1985ApJ...292..371D,1986ApJ...304...15B}.
Although these efforts ultimately proved fruitless, understanding of bias remains important.
In recent years much effort has been put into investigating galaxy
bias through theory and numerical modelling, while observational
results have been restricted by small survey volumes. With the advent
of large galaxy redshift surveys such as the 2dF Galaxy
Redshift Survey (2dFGRS: Colless et al. 2001, 2003)\nocite{2001MNRAS.328.1039C}, 
the Sloan Digital Sky Survey (SDSS: Strauss et al. 2002) 
\nocite{2002AJ....124.1810S}and the Deep Extragalactic Evolutionary Probe 
(DEEP: Davis et al. 2002) \nocite{2003SPIE.4834..161D}it is becoming possible
to quantify the galaxy distribution as never before, and provide
detailed descriptions with which to compare theoretical and numerical models.

In principle the form of bias should be derivable from the fundamental
physical processes involved in galaxy formation; until we
understand these, bias remains a description of our ignorance.
The simplest model of galaxy biasing is the linear biasing model: 
$\delta_g({\bf x}) = b\delta_m({\bf x})$ where $\delta_g$ is
the galaxy overdensity perturbation, $\delta_m$ the mass overdensity perturbation 
and $b$ a constant bias parameter. This model is unphysical for $b>1$, as
it allows negative densities. Alternative models in the
literature fall into several basic classes: linear or
non-linear, local or non-local, deterministic or stochastic.
Locally biased galaxy formation \citep[e.g.][]{1993MNRAS.262.1065C,1998ApJ...504..607S,1993ApJ...413..447F} 
depends only on the properties of the local environment, and the galaxy density is 
assumed to be a universal function of the matter density:
\be
\delta_g = f(\delta_m).
\ee
Because galaxies are discrete objects, this prescription is normally
supplemented by the {\it Poisson Clustering Hypothesis\/}, in 
which galaxies are modelled as random events, whose expectation 
number density is specified via $\delta_g$. This model for discreteness 
can only be an approximation, but there is no simple alternative.
We therefore assume Poisson sampling in what follows; for consistency,
theoretical predictions are treated in the same way as the real data.

Non-local models \citep[e.g.][]{1993ApJ...405..403B,1999ApJ...525..543M} 
arise when the efficiency of galaxy formation is
modulated over scales larger than those over which the matter moves,
for example by effects of quasar radiation on star formation. 
Stochastic bias (Pen 1998\nocite{1998ApJ...504..601P},
Dekel \& Lahav 1999\nocite{1999ApJ...520...24D}, hereafter DL99)
allows a range of values of $\delta_g$ for a given $\delta_m$, above the
Poissonian scatter caused by galaxy discreteness. 
Stochasticity is a natural part of non-local models, but some stochasticity is always expected
to arise from physical processes of galaxy formation
\citep{1999ApJ...522..590B}. 
Throughout this paper we follow the general framework for non-linear
stochastic biasing of DL99, in which  the overdensity
of one field can be related to that of a second field contained in the
same volume of space through
\be
\delta_1 = b(\delta_2) \delta_2 + \epsilon.
\ee
The scatter (stochasticity) in the relation is given by
\be
\epsilon \equiv \delta_1 - \langle\delta_1\rangle.
\ee
In principle the bias parameter $b$ can be any function of $\delta_2$;
a constant value of $b$
and $\epsilon=0$ represents deterministic linear biasing.

Galaxy bias is clearly of astrophysical interest in relation to
an understanding of galaxy formation and evolution.
Bias is also a major practical source of uncertainty in deriving cosmological
constraints from galaxy surveys. Some particular examples are the
measurement of $\beta = \Omega_m^{0.6} / b$ 
\citep{2001Natur.410..169P,2003MNRAS.346...78H}, where
DL99 showed that stochastic effects
could explain large discrepancies between results from different
methods \citep[for a review see][Table 7.2]{1999fsu..conf.....D}.
Power spectrum measurements require constant bias as a fundamental assumption
\citep{2001MNRAS.327.1297P}, and constraints placed on neutrino mass also assume scale
independent biasing \citep{2003JCAP...04..004E}. \citet{1998ApJ...504..601P} calculate the
effect of nonlinear stochastic bias on the measurement of the galaxy power spectrum on
large scales, showing how the galaxy variance, bias and galaxy-dark matter cross correlation
coefficient can be calculated from velocity distortions in the power
spectrum. The importance of biasing has increased still further with
the release of the WMAP first year results \citep{2003ApJS..148..175S}. In order to
combine CMB and 2dFGRS data to give tighter constraints on
cosmological parameters, a model for galaxy bias is required
\citep{2003ApJS..148..195V}. 

Three independent methods have been used to investigate
galaxy biasing in the 2dFGRS catalogue. \citet{2002MNRAS.333..961L}
combined pre-WMAP CMB and 2dFGRS
datasets to measure the average bias over a range $0.02 < k< 0.15\,
h\,{\rm Mpc}^{-1}$, concluding that galaxies are almost exactly unbiased on
these scales.  \citet{2002MNRAS.335..432V} found the bias parameter to be consistent
with unity over scales $0.1<k<0.5\,h\,{\rm Mpc}^{-1}$ through measurements of
the 2dFGRS bispectrum.

A more direct method of studying the relation between mass and light is to
map the dark matter using gravitational lensing. This field has made great
progress in recent years, and it has been possible to measure not only the
absolute degree of bias, but also its nonlinearity and stochasticity
\citep{2000AJ....120.1198F,2002ApJ...577..604H,2003ApJ...594...33F,2003MNRAS.346..994P}.
For example, \citet{2002ApJ...577..604H} combine the Red-Sequence Cluster
Survey and VIRMOS-DESCART survey to find an average bias $b=0.71$ and
linear correlation coefficient of $r \simeq 0.57$ on scales of
$1-2\mpcoh$. However, current weak lensing measurements are dominated by
non-linear and quasi-linear scales in the power spectrum, and it is not
yet possible to say a great deal about bias in the very large-scale linear
regime. This is of course the critical region for the interpretation of
redshift surveys, where we want to know the relation between the power
spectra of mass and light on $>100\Mpc$ scales.

This question will be settled by future weak lensing surveys. In the
meantime, we can address a related simpler problem: the {\it relative\/}
bias between subsets of galaxies. The morphological
differences between galaxies and the link to their environments has
been discussed for many decades as a potential clue to the nature and
evolution of galaxy clustering 
\citep[e.g.][]{1951ApJ...113..413S,1972ApJ...176....1G,1976ApJ...208...13D,2001ApJ...558..520Y}.
Modern galaxy redshift
surveys allow us to split the galaxy population into a variety of
subdivisions such as spectral type, colour and surface brightness. We
can look at relative bias as a function of scale, and weighted by
luminosity. This should yield important insights into the absolute degree
of bias that may exist. \citet{2001MNRAS.328...64N} measured bias as a function
of luminosity in the 2dFGRS, finding a bias relative to $L^*$
galaxies of  $b/b^* = 0.85+ 0.15L/L^*$.
We concentrate on the natural bimodality of the galaxy
population, between red early types with little active
star formation, and the blue late type population \citep[e.g.][]{2003IAUS..216E.201B}.
Lahav \& Saslaw (1992) measured bias as a function of morphological
type and scale using the UGC, ESO and IRAS catalogues. 
The Las Campanas Redshift
Survey has already provided some observational evidence against the linear
deterministic model from splitting galaxies by their spectral
types \citep{1999ApJ...518L..69T,2000ApJ...544...63B}, 
and we present here a more extensive analysis of this type.

There are several complementary methods for the measurement of galaxy
clustering, although most previous studies of the relative bias between galaxy types have
concentrated on a relative bias parameter defined as the square root
of the ratio of the correlation functions for the types under study.
\citet{2003MNRAS.344..847M} used this method to
measure the relative bias in the 2dFGRS, finding $b$(passive/active)
ranging from about 2.5 to 1.2 on scales $0.2\mpcoh < r < 20\mpcoh$.
However, even within such a large survey as the 2dFGRS the
correlation functions become noisy beyond
about $10\mpcoh$. A second method is counts-in-cells, which can be
directly related to the correlation function
\citep{1980lssu.book.....P}, and is
optimised for the study of larger scales. 
It is this latter method that we employ in this paper.
Conway et al. (2004) have also investigated the relative bias of
different galaxy types using a counts-in-cells analysis of the
2dFGRS, but they use magnitude limited samples, and consider only
deterministic bias models, whereas our present analyses use volume
limited samples, and consider stochastic bias models.  The
counts-in-cells method has also been used to calculate the variance and
higher order moments of galaxy clustering in the 2dFGRS (Conway et
al. 2004; Croton et al. 2004a,b; \citealt{astro-ph/0401405}).
\nocite{astro-ph/0404276,astro-ph/0401434,astro-ph/0401406} 

Many theoretical results on biasing from numerical models have also been reported. 
There are two main approaches to modelling galaxy distributions: semianalytic 
(e.g. \citealt*{1997MNRAS.286..795K}; \citealt{2000MNRAS.316..107B, 1999MNRAS.310.1087S})
and hydrodynamic
\citep[e.g.][]{1992ApJ...399L.113C,1999ApJ...522..590B,2000ApJ...538...83C,2001ApJ...558..520Y}.
Comparisons are given by \citet{2003MNRAS.338..913H} and \citet{2002MNRAS.335..762Y}.
Several studies have been made of galaxy biasing in these numerical
simulations \citep[e.g.][]{2001MNRAS.320..289S,2001ApJ...558..520Y},
but none provide results in sufficient detail to allow an easy
comparison with the 2dFGRS. We therefore analyse
a large new semianalytic calculation which is capable
of yielding mock results that can be analysed in an identical manner to the real data.

In this paper we concentrate on a few aspects of relative bias
mentioned above, splitting galaxies by spectral type and colour. We
investigate the nonlinearity, stochasticity and scale dependence of
the biasing relation through comparison with three models. Section 2
summarises the DL99 framework for biasing,
and presents the bias models used in this paper. Section 3
describes the 2dFGRS catalogue, the derivations of the galaxies
spectral types and colours and Section 4 explains the counts-in-cells
method. In Section 5 we show the methods used for model fitting and
error estimation. Section
6 gives the results and we compare our results with simulations in
Section 7.

Throughout, we adopt a cosmological geometry with $\Omega_m=0.3$,
$\Omega_v=0.7$ in order to convert redshifts and angles into three-dimensional
comoving distances. We define our cells with $h=0.7$, and all cell
lengths are quoted in $\Mpc$, instead of the standard $\mpcoh$.

\section{Modelling Relative Bias}\label{models}
The simplest model for any bias (i.e. mass-galaxy,
early-late, red-blue etc.) is that of {\it linear deterministic\/}
bias: given a number of one type of object you can
predict precisely (within Poisson errors) the number of the other type
of object in the same region of space, and the relationship between
the two numbers is linear. Recalling the familiar relation for the
mass/galaxy distributions $\delta_g = b\delta_m$, we can write
$\delta_\Lj = b\delta_\Ej$
where $\delta_\Ej$ ($\delta_\Lj$) denotes the
overdensity of early (late) type galaxies in a volume of space.
As described above, this empirical model 
can become unphysical in low density regions. 
Considering the complex processes involved in
galaxy formation, it would be surprising to find linear
deterministic biasing to be true in all cases. Any reasonable physical
theory in fact predicts non-trivial mass/galaxy biasing
\citep{1989MNRAS.237.1127C} and simulations can also find biasing 
to be a complicated issue particularly on small scales
\citep{1992ApJ...399L.113C,1999ApJ...522..590B,2001MNRAS.320..289S}. 

We investigate two potential improvements to the linear deterministic
model. Firstly the bias could be nonlinear, and some nonlinearity 
is inevitable in order to `fix' the unphysical properties of the linear model. 
Secondly, there may exist stochasticity (scatter beyond
Poissonian discreteness noise), due to astrophysical processes involved in galaxy
formation. DL99 presented a general framework to
quantify these different aspects of biasing, and the
following Section summarises their results.

\subsection{A framework for nonlinear, stochastic bias}\label{sec_frame}
We use the notation $f(\delta_\Ej)$ and $f(\delta_\Lj)$
to denote the one-point probability distribution functions (PDFs) for 
the fractional density fluctuations of early
and late type galaxies.
The fields $\delta_\Ej$ and $\delta_\Lj$
have zero mean by definition, and their variances are defined by 
\be 
\label{eq_sigma}
\sigma_i^2 \equiv \int_{-1}^{\infty} \delta_i^2 f(\delta_i)d(\delta_i)
\equiv \langle \delta_i^2 \rangle.
\ee
The joint underlying probability distribution of early and late type
galaxies is given by
\begin{eqnarray}
f(\delta_\Ej, \delta_\Lj) & = & f(\delta_\Ej)
f(\delta_\Lj|\delta_\Ej) \label{eq_prob1}\\
& = & f(\delta_\Lj) f(\delta_\Ej|\delta_{\Lj}).\label{eq_prob2}
\end{eqnarray}
Both equations (\ref{eq_prob1}) and (\ref{eq_prob2}) should give the
same outcome, but we have chosen to work with equation
(\ref{eq_prob1}) to avoid unphysical linear biasing.

The natural generalisation of linear biasing is given by
\be\label{eq_b}
b(\delta_\Ej)\delta_\Ej \equiv \langle\delta_\Lj|\delta_\Ej\rangle = \int f(\delta_\Lj|
\delta_\Ej)\delta_\Lj d\delta_\Lj.
\ee
There are several useful statistics that can be used to investigate
independently the fraction of nonlinearity and stochasticity of a model or
data. Firstly the mean biasing is defined by
\be\label{eq_hatb}
\hat{b} \equiv \frac{\langle b(\delta_\Ej) \delta_\Ej^2
\rangle }{\sigma_\Ej^2};
\ee
the nonlinear equivalent of this is
\be\label{eq_tildeb}
\tilde{b}^2 \equiv \frac{\langle b^2(\delta_\Ej) \delta_\Ej^2
\rangle }{\sigma_\Ej^2}.
\ee
In each case the denominator is assigned such that linear biasing
reduces to $b=\hat{b}=\tilde{b}$.
The {\it random biasing field\/} is defined as
\be
\mbox{\boldmath $\epsilon$} \equiv \delta_\Lj - \langle \delta_\Lj
| \delta_\Ej \rangle
\ee
and the statistical character of the biasing relation can be described via
its variance, the {\it biasing scatter function\/}
\be
\sigma_b^2(\delta_\Ej) \equiv \frac{\langle \mbox{\boldmath $\epsilon$}^2 |
\delta_\Ej \rangle}{\sigma_\Ej^2}.
\ee
The average biasing scatter is then given by
\be\label{eq_sigb}
\sigma_b^2 \equiv \frac{\langle \mbox{\boldmath $\epsilon$}^2 \rangle}{\sigma_\Ej^2}.
\ee

The purpose of this parameterisation is to separate naturally the
effects of nonlinearity and stochasticity, allowing them to be
quantified via the relations 
\be
{\bf nonlinearity}  \equiv  \tilde{b}/\hat{b}\label{eq_nonlin}
\ee
\be
{\bf stochasticity}  \equiv  \sigma_b/\hat{b}. \label{eq_stoc}
\ee

There are two further useful relations that are often quoted in the
literature as measures of the bias parameter and stochasticity: the
ratio of variances
\be\label{eq_bvar}
b_{\rm var} \equiv \frac{\sigma_\Lj} {\sigma_\Ej}
\ee
and the linear correlation coefficient
\be\label{eq_r}
r_{\rm lin} \equiv \frac{\langle \delta_\Ej \delta_\Lj \rangle}
{ \sigma_\Ej \sigma_\Lj} = \frac{\hat{b}}{b_{\rm var}}.
\ee
Both $b_{\rm var}$ and $r_{\rm lin}$ can be written in terms of the basic
parameters given above, and both mix nonlinear and stochastic
effects. 
Non-parametric correlation coefficients can also be calculated directly from
the data, although some method must be employed to account for shot
noise. 
We refer the interested reader to DL99 for further details on the equations
in this Section.

Note that we work throughout with redshift-space
overdensities. Redshift-space distortions are dependent on galaxy type    
due to the different clustering properties of early and late type 
galaxies. On nonlinear scales the dominant effect is the finger-of-god stretching,
but on the scales of interest to this paper we expect the
linear $\beta$-effect to apply \citep{1987MNRAS.227....1K}. Averaging
over all angles and including stochasticity between the galaxy and
matter fields, we can write the 
redshift-space power spectrum, $P_{\rm s}$, as
\be
\frac{P_{\rm s}}{P_{\rm r}} = (1 + \frac{2}{3}r\beta + \frac{1}{5}\beta^2)
\ee
where $\beta = \Omega_m^{0.6}/b$, $b$ is the mass-galaxy bias, $r$ is
the linear mass-galaxy correlation coefficient and
$P_{\rm r}$ is the real-space power spectrum \citep{1998ApJ...504..601P,1999ApJ...520...24D}. The
$\beta$-effect was measured for galaxies of different spectral class
in the 2dFGRS by  \citet{2003MNRAS.344..847M}, obtaining $\beta_\Lj =
0.49 \pm 0.13$ and $\beta_\Ej = 0.48 \pm 0.14$. From these results and
assuming $r=1$ we obtain 
\be
\frac{P_{\rm s,E}}{P_{\rm s,L}} = 0.99\frac{P_{\rm r,E}}{P_{\rm r,L}}.
\ee
Although this suggests the effect is not large and currently
insignificant within the errors, it is clear that in the case of zero
stochasticity, redshift-space distortions will work to reduce the
difference in the measured clustering between types. However, including
a value of $r$ which is non-unity and dependent on galaxy type as
suggested by simulations \citep[e.g.][]{1999ApJ...522..590B}, has a
significant effect. For example, taking $r_\Lj=0.8$ and $r_\Ej=1.0$ increases the
relative distortion from 0.99 to 1.04, where $r_\Lj$ ($r_\Ej$) is the
linear correlation coefficient between the mass and late (early) type galaxy
fields.


\subsection{One point probability distribution function}\label{pdf_one}
Given equation (\ref{eq_prob1})
we can split the model into two parts, firstly the distribution
of early type galaxies per cell, and secondly the biasing relation
connecting the two distributions (see the following Section).
A standard description for the underlying probability distribution of
a galaxy overdensity, $f(1+\delta)$, is lognormal
\citep{1991MNRAS.248....1C}.
Applying this for example to the early type galaxies:
\be\label{eq_onepdf}
f(\delta_\Ej) \, d\delta_\Ej = 
\frac{1}{\omega_\Ej\sqrt{2\pi}}
\exp\left[{\displaystyle \frac{-x^2}{2\omega_\Ej^2}} \right] \; dx
\ee
where
\be
x = \ln(1+\delta_\Ej) + \frac{\omega_\Ej^2}{2}
\ee
and $\omega_\Ej^2$ is the variance of the corresponding normal
distribution $f[\ln(1+\delta)]$:
\be
\omega_\Ej^2 = \langle[\ln(1+\delta_\Ej)]^2\rangle.
\ee
The offset $\omega_\Ej^2/2$ is required to impose $\langle \delta_\Ej \rangle=0$.
If the lognormal distribution correctly describes the data,
the variance of the overdensities, $\langle \delta_\Ej^2 \rangle$, is related to the variance of
the underlying Gaussian distribution by 
\be
\sigma_\Ej^2 \equiv \langle \delta_\Ej^2 \rangle = \exp[\omega_\Ej^2]-1.
\ee
In Section \ref{efst90} we show that fitting a lognormal distribution directly to the data does not
yield quite the same values for $\sigma_\Ej$ and $\sigma_\Lj$ as a
direct variance estimate, but this does not affect our final
results for stochasticity.
On the largest scales, a lognormal distribution is completely
consistent with the 2dFGRS data, and it provides a transparent and simple
way to describe the density field.

\subsection{Biasing models}
\subsubsection{Deterministic bias: linear and power law}
Firstly concentrating on deterministic bias, we can write the joint
probability distribution function as
\be\label{eq_det}
f(\delta_\Lj|\delta_\Ej) = \delta^{\rm D}(\delta_\Lj - b(\delta_\Ej)\delta_\Ej)
\ee
where $\delta^{\rm D}$ is the Dirac delta function. 
This reduces directly to linear bias  by setting 
\be\label{eq_lin}
b(\delta_\Ej)\delta_\Ej=b_{0,\mathcal L} + b_{1,\mathcal
L}\delta_\Ej
\ee
where the constraint $\langle\delta_\Lj\rangle = 0$ fixes $b_{0,\mathcal L} = 0$.
A simple variation could be power law bias
\be\label{eq_plaw}
b(\delta_\Ej)\delta_\Ej =
b_{0,\mathcal P}(1+\delta_\Ej)^{b_{1,\mathcal P}} -1
\ee
which avoids the negative density
predictions of linear bias, and reduces to the linear
biasing relation near $\delta=0$.  Rearranging equation
(\ref{eq_plaw}), using the properties of lognormal distributions and
the fact that $b_{1,\mathcal P} = \omega_\Lj / \omega_\Ej$, we find
for power law bias that 
\be
b_{0,\mathcal P} = \exp[0.5\omega_\Ej^2(b_{1,\mathcal P}
-b_{1,\mathcal P}^2)].
\ee
For convenience we define 
\be
b_{\rm lin} = b_{1,\mathcal L}
\ee
and 
\be
b_{\rm  pow} = b_{1,\mathcal P}
\ee
throughout the rest of this work.

\subsubsection{Stochastic bias: bivariate lognormal}
Returning to equation (\ref{eq_b}), we can introduce a broader function
than the Dirac delta function of equation (\ref{eq_det}). 
An interesting class of model is when both $\delta$ fields form a
bivariate Gaussian distribution, but this again becomes unphysical for
$\delta< -1$ (DL99). It is however simple to 
cure this defect by assuming instead
a {\it bivariate lognormal\/} distribution, for which the joint probability
distribution is given by
\be
f(g_\Ej, g_\Lj)  = 
{|V|^{-1/2}\over 2\pi}
\exp\left[- \frac{ (\tilde{g_\Ej}^2 +\tilde{g_\Lj}^2 - 
2r_{\LN}\, \tilde{g_\Ej} \, \tilde{g_\Lj})} 
{2(1-r_{\LN}^2)}\right],
\ee
where $g_i = \ln(1+\delta_i) - \langle \ln(1+\delta_i)\rangle$ and $\tilde{g_i} =
g_i/\omega_i$, with $i$ corresponding to early or late type.
$\omega_i$ is related to the variance of the underlying Gaussian
field $\ln(1+\delta_i)$ as for the one--point lognormal distribution:
\be 
\sigma_i^2 \equiv \langle \delta_i^2 \rangle = \exp[\omega_i^2]-1.
\ee
The correlation coefficient is
\be\label{eq_rln}
r_{\LN} = \frac{\langle g_\Ej g_\Lj
  \rangle}{\omega_\Ej \omega_\Lj}
\equiv \frac{\omega_{\japsub EL}^2}{\omega_\Ej \omega_\Lj}
\ee
and $V$ is the covariance matrix
\be\label{eq_V}
  V = \left(
    \begin{array}{lc}
    \omega_\Ej^2  & \omega_{\japsub EL}^2  \vspace{0.1cm}\\
    \omega_{\japsub EL}^2 & \omega_\Lj^2 
    \end{array}
  \right).
\ee

Taking $f[\ln(1+\delta_\Ej)]$ to be a Gaussian of width
$\omega_\Ej$
and mean $-\omega_\Ej^2/2 $ (i.e. $f(\delta_\Ej)$ is distributed as a
lognormal, equation [\ref{eq_onepdf}]), the conditional
probability distribution is 
\begin{eqnarray}
f(g_\Lj| g_\Ej) & = & 
\frac{f(g_\Ej, g_\Lj)}{f(g_\Ej)}\nonumber\\
& = & \frac{\omega_\Ej}{(2\pi|V|)^{1/2}} 
\exp\left[-\frac{(\tilde{g_\Lj}-r_\LN\,\tilde{g_{\Ej}})^2}{2(1-r_\LN^2)}\right], \label{eq_bivLNcond}
\end{eqnarray}
i.e. the distribution of $\tilde{g_\Lj}|\tilde{g_\Ej}$ is a Gaussian with mean $r_{\LN}
\,\tilde{g_\Ej}$ and variance $1-r_{\LN}^2$. 

As $r_{\LN} \rightarrow 1$, equation (\ref{eq_bivLNcond})
reduces to a Dirac delta function, and
this bivariate lognormal model reduces to the power law bias model of
equation (\ref{eq_plaw}). It is important to note that $r_{\LN}$ is not
equal to the linear correlation coefficient $r_{\rm lin}$ of equation
(\ref{eq_r}), which can differ from unity even if $r_{\LN}=1$.
In this sense, the lognormal parameters offer a cleaner separation
of stochastic and nonlinear effects.
If stochasticity is present within the data, this model may
provide an improvement over the deterministic biasing models.
As observational data improve, it may become possible to constrain
the relative biasing function to a greater extent; the current
data are insufficient for such an analysis. 

Analytic solutions exist to the mean biasing parameters and biasing
scatter function given in Section \ref{sec_frame} for this bivariate
lognormal model. These relations are presented in Appendix \ref{app1}.

\subsection{Including observational shot noise}
It is not possible to measure the underlying probability
distribution $f(\delta_\Ej,\delta_\Lj)$ directly due to contamination of the
observational data with noise, the dominant form of which is expected
to be Poisson or `shot' noise. This can be included in the models of
the previous Section by convolution with a Poisson distribution
\citep{1991MNRAS.248....1C,2000ApJ...544...63B}.
In this way the measured probability of finding $N_\Ej$ early type
galaxies and $N_\Lj$ late type galaxies within a cell,
$P(N_\Ej,N_\Lj)$, can be compared with the models
above. Accounting for shot noise in this way results in the
models being less sensitive to outliers than equations
(\ref{eq_det}) and (\ref{eq_bivLNcond}).

Using equation (\ref{eq_prob1}) to combine the one point PDF
(\ref{eq_onepdf}) with the conditional PDF (\ref{eq_det} or
\ref{eq_bivLNcond}), provides a model for the actual joint probability
distribution function $f(\delta_\Ej,\delta_{\Lj})$. Convolution with a Poisson distribution then gives 
\begin{eqnarray}\label{eq_2d}
\lefteqn{P(N_\Ej, N_\Lj)  =  \int_{-1}^{\infty}
\int_{-1}^{\infty} \frac{\bar{N}_\Ej^{N_\Ej}(1+\delta_\Ej)^{N_\Ej}}
{N_\Ej!}e^{-\bar{N}_\Ej(1+\delta_\Ej)} f(\delta_\Ej)} \nonumber \hspace{0.4cm}\\
& &\times \frac{\bar{N}_\Lj^{N_\Lj}(1+\delta_\Lj)^{N_\Lj}}
{N_\Lj!}e^{-\bar{N}_\Lj(1+\delta_\Lj)}f(\delta_\Lj|\delta_\Ej)d\delta_\Ej d\delta_\Lj,
\end{eqnarray}
where $\bar{N}_\Ej$ ($\bar{N}_\Lj$) is the expected number of early (late)
type galaxies in a given cell, allowing for completeness.

\section{The Data: the 2\lowercase{d}FGRS}
The 2dF Galaxy Redshift Survey (2dFGRS), carried out between May 1997 and April
2002, has obtained 221,414 good quality galaxy spectra using the multi-object
spectrograph 2dF on the Anglo-Australian Telescope. The main survey
area comprises two rectangular strips of sky with boundaries 
$(09^h50^m < \alpha < 14^h50^m, -7.5\deg < \delta <+2.5\deg)$ for the
NGP and $(21^h40^m < \alpha < 03^h40^m, -37.5\deg < \delta <
-22.5\deg)$ for the SGP, with a galaxy median redshift of $\bar z =
0.11$. At the median redshift, the physical size of the survey strips
is $375\mpcoh$ long and the SGP and NGP regions
have widths of $75\mpcoh$ and $37.5\mpcoh$ respectively
\citep{2003AIPC..666..275P}.
The input galaxies were selected from a revised and extended
version of the APM galaxy catalogue \citep{1990MNRAS.243..692M}, and have a
limiting extinction corrected magnitude of $\bj = 19.45$. Further
details of the 2dFGRS can be found in
\citet{2001MNRAS.328.1039C}, \citet{astro-ph/0306581} and on the web at
{\tt http://msowww.anu.edu.au/2dFGRS/}.

For any structure analysis it is
important to be aware of several problems that cause varying
completeness over the survey region. Common to
all similar surveys, some regions of the sky must be masked due to
bright stars causing internal holes. Furthermore, due to the
adaptive tiling algorithm employed to ensure an optimal observing
strategy, the sampling fraction falls to as little as 50\% near the
survey boundaries and internal holes due to lack of tile
overlap. Subsequent reanalysis of the photometry of the APM galaxy
catalogue has shown the survey depth to vary slightly with position
on the sky. To account for this, we use a limiting corrected magnitude
of $\bj=19.2$.

\begin{figure*}
\begin{minipage}[t]{15cm}
\begin{minipage}{7.5cm}
\includegraphics[scale=0.4]{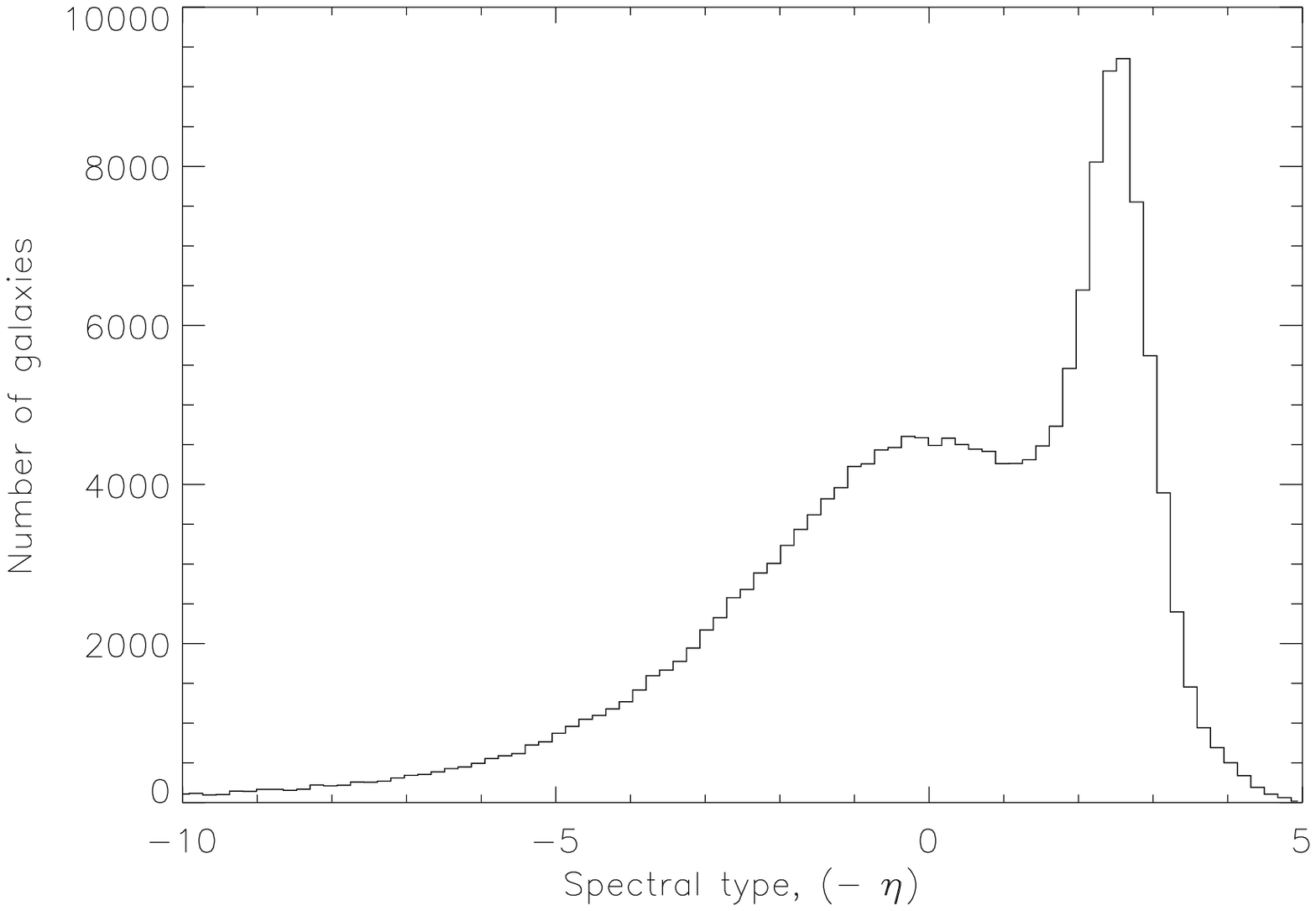}
\end{minipage}
\begin{minipage}{7.5cm}
\includegraphics[scale=0.4]{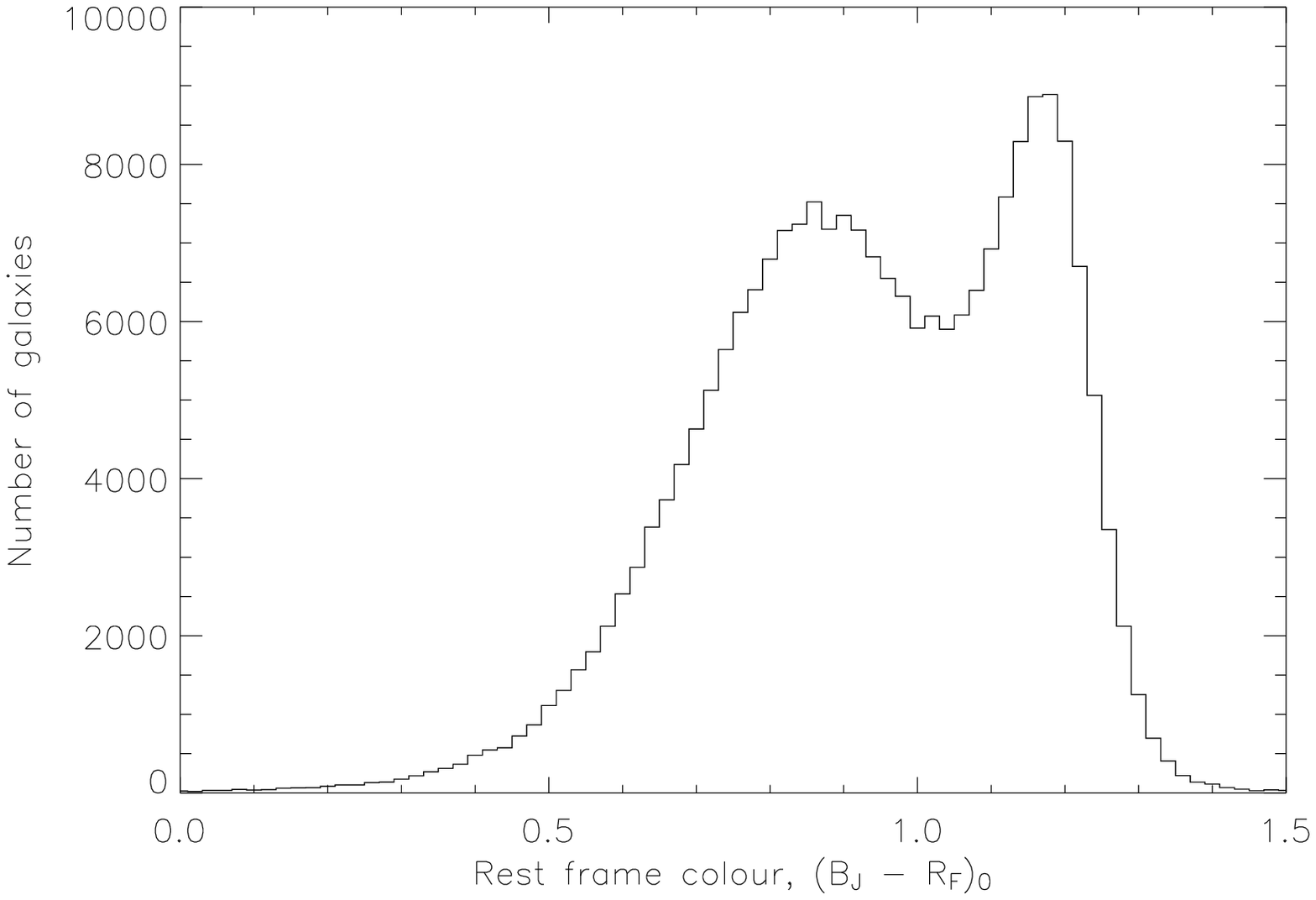}
\end{minipage}
\caption{\small The distributions of spectral type and rest frame colour for all
  2dFGRS galaxies. The distinction between passive and actively
  star-forming galaxies is clear in both distributions. Cuts at
  $\eta = -1.4$ and $(B-R)_0 = 1.07$ produce the four subgroups with
  which we work.}
\label{fig_etacol}
\end{minipage}
\end{figure*}

\subsection{The volume limited galaxy sample}\label{sec_galsampl}
It is well known that the luminosity of a galaxy is correlated with
galaxy type. Therefore in a flux limited sample the fraction of
early/late types varies with redshift,
potentially complicating the analysis. Within any redshift
survey the number density of objects drops substantially as we reach
beyond the $L_*$ galaxy luminosity. The size of the
2dFGRS presents the option of studying volume limited
galaxy samples rather than the more usual flux limited datasets of
previous galaxy redshift surveys. By imposing a
luminosity and redshift cut, volume limited samples contain a
representative sample of most galaxies over a large redshift
range. Although some faint galaxies at low redshift are lost from the
analysis, the sample selection effects are greatly simplified.

We use the publicly released data of June 2003, containing a total of
221,414 unique galaxies with reliable redshifts,
192,979 of which have spectral classification.
An absolute magnitude limit of $M_{\bj}-5\log_{10}(h) \leq -19.0$ gives a
representative sample of the local population, maximising the number
of cells, versus the number of galaxies in each cell. The absolute
magnitude is given by $M_{\bj} = m - DM -K(z)$, where m is the
apparent extinction corrected magnitude, $DM$ is the
distance modulus, and $K(z)$ the K-correction. Setting a
limiting survey magnitude of $m=19.2$ allows for the varying survey
depth with position on the sky \citep{2001MNRAS.328.1039C}, the
K-corrections as a function of
$\eta$ type are given in \citet{2002MNRAS.333..133M}.  
This gives a maximum redshift for our sample set by Type 1 galaxies of
$z_{\rm max} = 0.114$, and 48,066 galaxies in total. 46,912 of
these have a spectral classification and 48,040 have measured colours. 
To account for the selection effects within the survey we use the
publicly available redshift completeness masks. These are sufficient
for galaxies classified by colour, but the
spectral type analysis introduces extra selection effects. A
difference in completeness over a region of sky could occur, for example,
when the spectra of a survey plate are of too poor quality to perform the
spectral type analysis, yet redshifts can be obtained. It is necessary
to create separate masks to include these effects using the publicly
available software.

\subsection{Galaxy properties in the 2dFGRS}
The 2dFGRS catalogue provides two methods of classification for
comparison. Firstly the well studied galaxy spectral type $\eta$,
and secondly the photometric colours of the galaxies have recently been
derived. 

\begin{figure}
\vspace*{-0.4cm}
\begin{minipage}{8.5cm}
\hspace*{-0.5cm}
\includegraphics[scale=0.5]{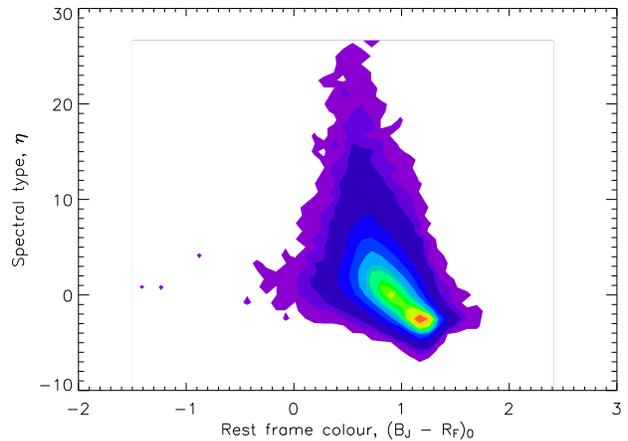}
\end{minipage}
\caption{\small The joint distribution of the colour and spectral types
  for galaxies in the 2dFGRS. 
}
\label{fig_jointetacol}

\end{figure}

\subsubsection{Spectral type, $\eta$}\label{sec_eta}
The spectral type of the galaxies has been derived by a Principal Component
Analysis (PCA), which identifies the most variable aspects of the
galaxy spectra with no prior assumptions or template spectra 
\citep{1999MNRAS.308..459F,2002MNRAS.333..133M}.
The spectral type of the 2dFGRS galaxies is characterised by the value
$\eta$, a linear combination of the first two principal components, 
derived in order to minimise the effect of distortions and imperfections in
the 2dFGRS spectra. 
In effect, $\eta$ classifies galaxies according to the average
emission and absorption line strength in the spectrum. 
$\eta$ provides a continuous
classification scheme, but for our purposes it is necessary to split
the galaxies into two classes, at $\eta=-1.4$ as suggested in
\citet{2003MNRAS.338..197M}. 
Galaxies with $\eta<-1.4$ (Type 1) are shown to be predominantly
passive galaxies and those with $\eta>-1.4$ (Types 2--4) predominantly
star-forming \citep{2003MNRAS.343..871M}. The former are hence termed
`early type' and the latter `late type'. The 2dFGRS catalogue
contains 74,548 early type and 118,424 late type galaxies defined in
this way.

One concern with using optical fibre spectra for
this type of analysis are `aperture effects', resulting from the
fixed aperture of the fibres being smaller than the size of galaxies.
This could result in, for example, only the bulge components of close spirals being
observed. Such effects have been studied in detail by \citet{2002MNRAS.333..133M}, and no systematic
bias found. One possible explanation for this is the poor
seeing present at the Anglo-Australian Telescope, of order
$1.5''$--$1.8''$, which will cause the fibre to average over a large fraction of 
the total galaxy light in most cases and dilute aperture effects. An overabundance
of late type galaxies
was detected at redshifts beyond 0.11, which could be attributed to
either aperture effects or evolution; however, this will not affect our
volume limited galaxy sample with a maximum redshift of $z_{\rm max}=0.114$. 
Aperture effects are discussed further in Section \ref{sec_zdep}.

\begin{table}
\centering
\caption{\small \label{tab0} Numbers of early/late and red/blue
  galaxies in the 2dFGRS catalogue with good quality spectra ($Q >= 3$). }
\vspace{0.2cm}
\begin{tabular}[b]{cccc} \hline \hline
& red/early & blue/late & Total\\
\hline
colour &77,120 &144,292 &221,414\\
$\eta$  &74,548  &118,424  &192,979\\
\end{tabular}
\end{table}

\subsubsection{Broad-band colours}
More recently it has been possible to obtain broad-band colours for
the 2dFGRS galaxies using the same $\bj$ UKST plates as the
survey input catalogue \citep{2001MNRAS.326.1279H}, but now scanned
with the SuperCosmos machine to yield smaller
errors of about 0.09 mag per band. Similar scans have also been made
of the UKST $R_F$ plates. The extinction corrections are from
the dust maps of \citet*{1998ApJ...500..525S} and wavelength dependent
extinction ratios are from \citet*{1989ApJ...345..245C}. 
We define rest frame colour 
\be
(B-R)_0 \equiv \bj - \rf - K(\bj) + K(\rf)
\ee
where the colour-dependent K corrections are 
\be
\eqalign{
K(\bj)  =&  (-1.63+4.53\times C)\times y \cr
         & + (-4.03-2.01\times C)\times y^2 \cr
         & - \frac{z}{(1+(10\, z)^4)}, \cr
}
\ee
\be
\eqalign{
K(\rf)  =&  (-0.08+1.45\times C)\times y \cr
         & +(-2.88-0.48\times C)\times y^2,\cr
}
\ee
with $y=z/(1+z)$ and $C = \bj - \rf$. 
See \citet{2004MNRAS.349..576C} for more details. A division at $(B-R)_0 = 1.07$ achieves a similar separation
between `passive' and `actively star forming' galaxies 
to spectral classification of Type 1 to Types 2--4, giving a total of 77,120 red galaxies and
144,292 blue galaxies.

The distributions of $\eta$ type and colour for the 2dFGRS galaxies
are shown in Fig. \ref{fig_etacol}, and the joint distribution is
shown in Fig. \ref{fig_jointetacol}. The correlation between the two
properties is clear, together with the distinct bimodality, yet it is
obvious that the relationship is not exactly one-to-one. Table \ref{tab0} gives the
respective numbers of each galaxy type in the 2dFGRS catalogue for comparison.

\section{Method: Counts-in-Cells}

\begin{figure*}
\begin{minipage}[tb]{17cm}
\hspace*{1cm}
\begin{minipage}{8.0cm}
\includegraphics[scale=0.55]{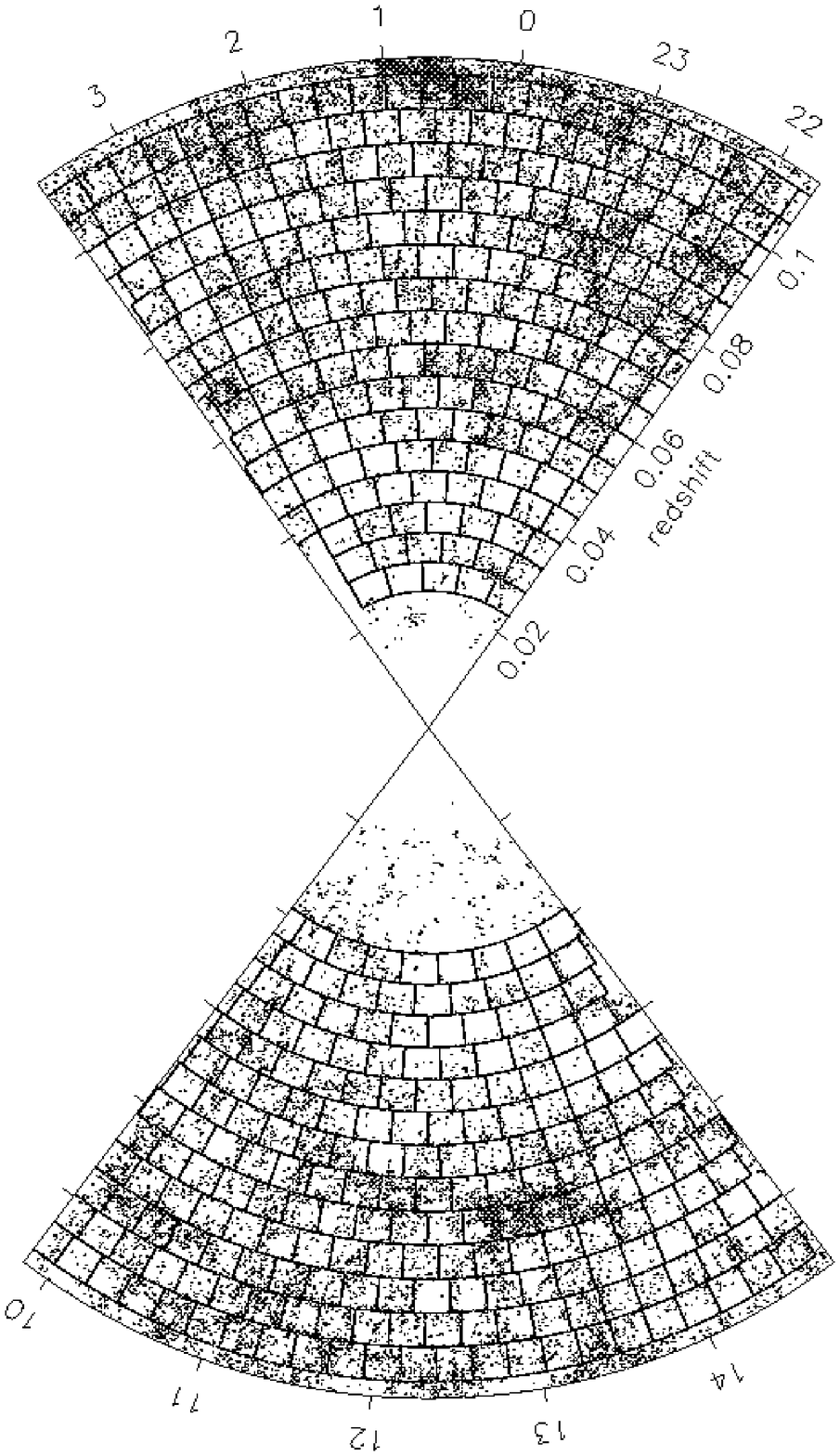}
\end{minipage}
\begin{minipage}{8.0cm}
\includegraphics[scale=0.55]{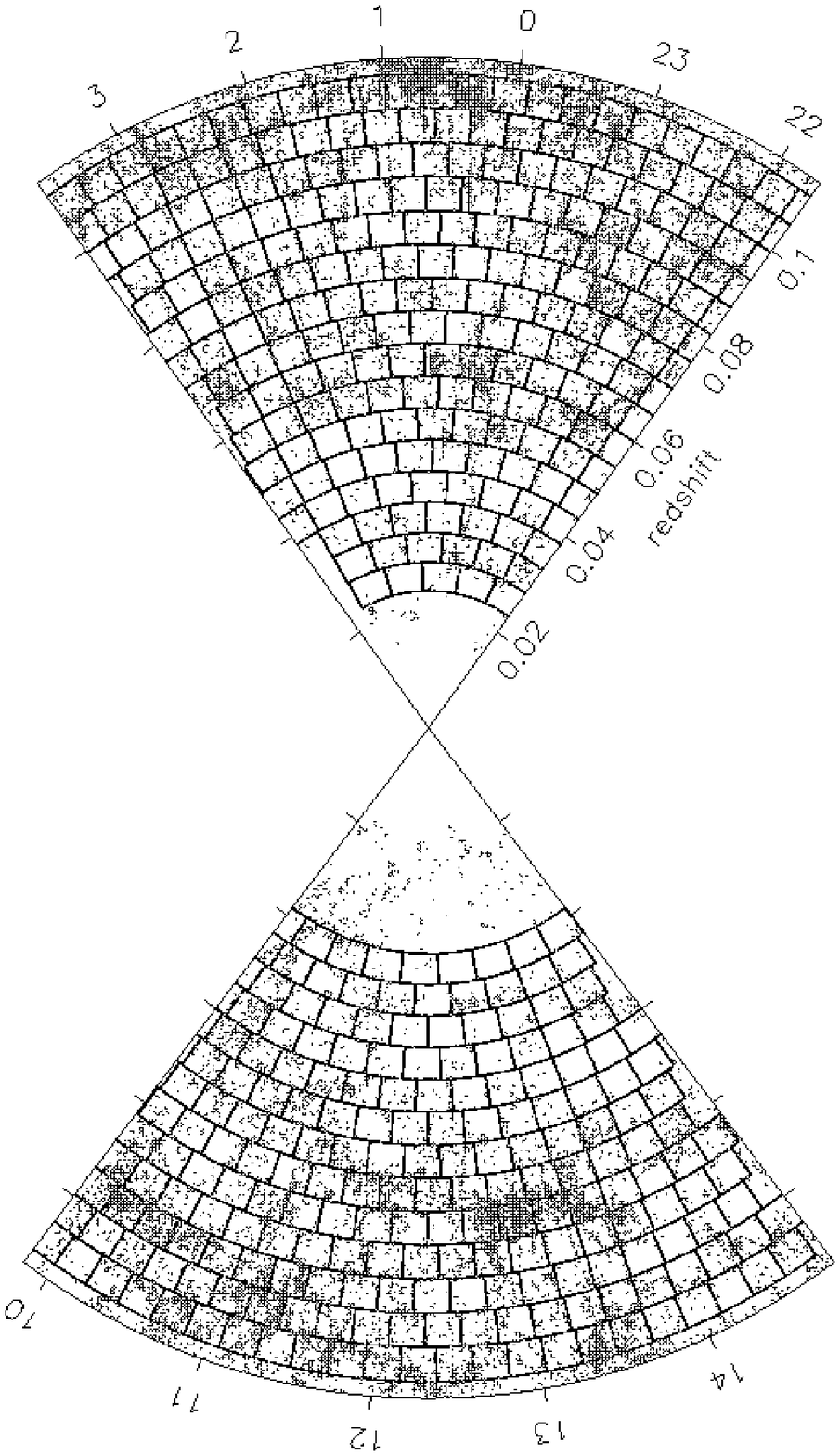}
\end{minipage}
\caption{\small Wedge plots of the 2dFGRS volume limited survey
  region with $M_{\bj}-5\log_{10}(h) \leq -19.0$. 
  Dots represent late type galaxies on the left and early type
  galaxies on the right (classified by spectral type).
  Redshift increases from the centre, and right ascension is shown on
  the horizontal axis, declination is projected onto the plane.
  Typical cell boundaries of length $L=25\Mpc$ are overplotted.
  }
\label{fig_boxes}
\end{minipage}
\end{figure*}

A counts-in-cells analysis is employed, which involves splitting the
survey region into a lattice of roughly cubical cells
and counting the number of galaxies in each cell. The cell dimensions
are defined such that all have equal volume
$V\equiv L^3$, but with limits to right ascension and declination that form a
square on the sky. This angular selection simplifies the treatment of the survey mask, 
but it means that the cells are not perfect cubes.
Over the redshift range involved, this effect is small.
The cells are required to fit strictly within the 2dFGRS
area, causing some parts of the survey to be unused.
Although this restriction in principle
removes any boundary effects, it means that cells of different sizes
sample slightly different areas of the universe. We define our cells with $h=0.7$ and in what
follows all cell lengths are quoted in Mpc, instead of the standard $\mpcoh$.
In practice, we considered
$10\Mpc \leq L \leq 45\Mpc$, giving a total of between 11,423 and
72 cells in the volume limited survey area after removing low
completeness cells. These cell sizes are
equivalent in volume to using top hat smoothing spheres with radii $6.1\mpcoh
\leq r \leq 27.9\mpcoh$. Fig.
\ref{fig_boxes} shows an example of how $25\Mpc$ cells cover the 2dFGRS
volume to $z=0.11$. 

Due to internal holes in the survey and the adaptive tiling algorithm
employed, the sampling fraction in the 2dFGRS varies over the
sky. Random 2dFGRS catalogues can be created, which include these selection
effects by making use of the calculated survey masks. Each cell count
is weighted by the fraction of random points found in the same cell in
the mock catalogue. The spectral type analysis
introduces extra selection effects, which are quantified by a special
mask (see Section \ref{sec_galsampl}).

An overdensity $\delta$ is calculated
for each cell by dividing the observed cell counts $N$ by the expected
number for a given cell allowing for completeness, $\bar{N}$: 
\be\label{eq_di}
\delta_i = \frac{N_i}{\bar{N}} - 1.
\ee
This procedure is carried out for both early and late type
galaxies within each cell. The overall density variance is defined by
equation (\ref{eq_sigma}).

\begin{figure}
\includegraphics[scale=0.55]{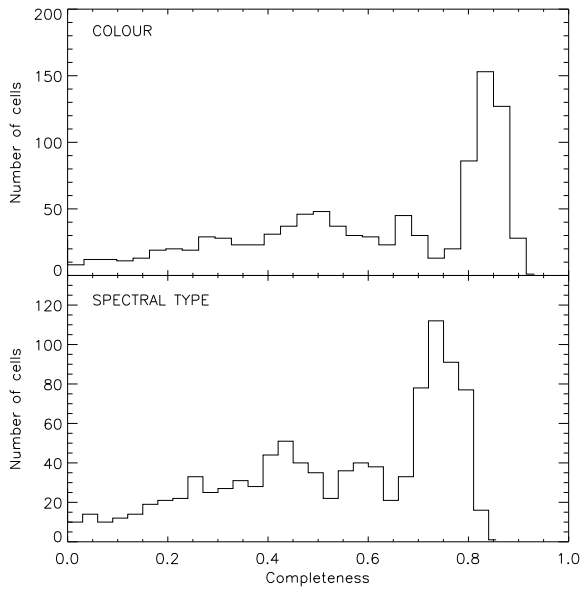}
\caption{\small Histograms showing the completeness distribution of
  $25\Mpc$ cells using the standard 2dFGRS mask (top),
  and including the effects of $\eta$ classification (bottom). }
\label{fig_compl}
\end{figure}

It is necessary to set a completeness limit 
to remove excessively under-sampled cells, such as
those affected by holes in the survey due to stars, or cells at the
less observed edges of the survey
volume. Although the limits are somewhat arbitrary, it is important
they are set correctly as incomplete cells could
affect our measurements of scatter in the biasing relation. 
Fig. \ref{fig_compl} shows the completeness distributions for
$L=25\Mpc$ cells. It can be seen that for both $\eta$ and colour
classification the
distribution has a sharp peak of almost complete cells, with a long
tail to low completeness and a sharp cut off at high completeness. 
The figure also highlights the importance of including the effects of $\eta$
classification on the 2dFGRS mask as the completeness peak and cut
off is noticeably lower for $\eta$ classification than for all galaxies.
The lower completeness is reflected in the reduced number of cells
available for analysis.

A completeness limit is set for each cell at 70\% (or 60\% for the
larger cells), to include all cells within the high completeness
peak. In order to check the effects of completeness on the final results,
the models were
also fit to only those cells with completenesses higher than 80\%
(70\% for the larger cells), and the results found to be consistent
within the errors. 

A general concern with a counts-in-cells analysis of observational
data is the varying survey selection function over a cell's extent.
For example, a cell containing a cluster of galaxies at
its most distant edge, and weighted by the average selection
function over its volume, would give a different `count' to a cell
containing a cluster near its inner boundary. Furthermore, with a
joint counts-in-cells analysis any relationship between luminosity and
galaxy type or colour will cause differing fractions
of objects with redshift. 

With careful use of type-dependent selection functions such effects
   can be allowed for (Conway et al. 2004), but in our analysis we
   use volume limited samples. This approach avoids these
   complications at the expense of reducing the number of galaxies in
   the analysis. The size of the 2dFGRS offers a great advantage over
   previous observational studies of relative bias, because it
   provides volume limited samples large enough to produce reliable
   measurements.

\section{Parameter fitting}
\subsection{A maximum likelihood approach}\label{sec_like}
Once we have chosen a model, a maximum likelihood method is used to
fit the free parameters of the model to the data.
Denoting the number of early (late) type galaxies within cell $i$ as
$N_{{\Ej},i}$ ($N_{\Lj,i}$), the likelihood of finding a
cell containing $N_\Ej$ early
type and $N_\Lj$ late type galaxies given a model with free parameters
$\bm{\alpha}$, is defined as
\be
L_i(N_{{\Ej},i},N_{{\Lj},i};\bm{\alpha})= P(N_{{\Ej},i},N_{{\Lj},i}|\bm{\alpha})
\ee
and the total likelihood for all cells is then 
\be
L = \prod L_i.
\ee
The likelihood can be maximised with respect to the free parameters
$\bm{\alpha}$ to find the best fitting values $\hat{\bm{\alpha}}$ for
the model given the dataset. In practice it is easier to minimise the function 
\be
\mathcal{L} \equiv -\sum_i \ln L_i.
\ee
Note that this definition of $\cal L$ differs by a factor of two
compared to Conway et al. (2004).

The models in Section \ref{models} contain two or three free
parameters: $\sigma_\Ej$ and/or $\sigma_\Lj$ from the one point PDFs, and $b$ or
$r_{\LN}$ from the conditional probability function. These parameters were
fitted simultaneously to the data using a downhill simplex method \citep{1992nrca.book.....P}.

\subsection{Error estimation}
As it is not possible to derive analytic solutions to the sampling
distribution of our maximum likelihood estimators $\hat{\bm{\alpha}}$,
the standard error on our parameters must be estimated directly from
the likelihood function using Bayes' theorem and assuming a uniform
prior on $\bm{\alpha}$:
\be
P(\bm{\alpha} | \bm{x}) \propto L(\bm{x};\bm{\alpha})
\ee
where $\bm{\alpha}$ again denotes the model parameters, $\bm{x}$
the data, and $P$ the probability.

For a single free parameter, the upper and lower limits on $\alpha$
are found from
\be\label{eq_err}
P(\alpha_- \le \alpha < \alpha_+|\bm{x}) = \frac{\int_{\alpha_-}^{\alpha_+}
L(\bm{x};\alpha)d\alpha }{\int_{-\infty}^{\infty} L(\bm{x};\alpha)d\alpha}
\ee
If it can be assumed that the likelihood function is reasonably
approximated by a Gaussian, $1\sigma$ errors on the
parameter can be estimated. 
For multi-parameter models it is necessary to quantify any possible
degeneracy between errors. If the multi-dimensional likelihood
function can be approximated by a multivariate Gaussian distribution,
individual errors and correlations between the parameters can be found.
  
A second method of error estimation involves creating many mock datasets
from the fitted model probability distributions themselves. These
datasets are made through Monte Carlo techniques, and designed to closely reproduce
the true data in size. On applying the above likelihood
techniques to these mock datasets, the best fit and true parameters
can be compared to estimate the errors. The advantage
of this method is that no assumptions need to be made about the shape of
the likelihood function. The disadvantage is that we are assuming that
the model is a correct representation of the data, as the errors
strictly apply only to the model not the data. By increasing
the size of the mock datasets, this method can also be used to check
for any bias inherent in the fitting method. This process was carried
out for each model in this paper, finding the parameter estimations to
be unbiased. 

In all cases, we will make the assumption that the density fluctuations
in each cell can be treated as independent. This is clearly not true
in detail, since the existence of modes with wavelength $\gsim L$ will cause
a correlation between nearby cells. This was considered by 
\citet*{1995ApJ...438...49B}, who showed that
the correlation coefficient was low even for adjacent cells: $r\simeq 0.2$.
As we shall see, it is $(1-r^2)^{1/2}$ that matters for joint distributions,
and so the failure of independence is negligible in practice.

\subsection{Model comparison}\label{sec_modelcomp}

Once we have found the best fit parameters for each of our three
models, we would like to know the goodness-of-fit of the models and the
significance of any differences between the fits. 
We approach this using two different methods.

\subsubsection{Likelihood ratios}
To test the significance of one model against
another model we use the likelihood ratio test. In its simplest form we
define the maximum likelihood ratio for hypothesis $H_0$ versus $H_1$
\be\label{eq_likeratio}
\lambda = \frac{L(\bm{x}|H_0)}{L(\bm{x}|H_1)}
\ee
where $\bm{x}$ is the data, and $L$ represents the {\it maximum\/}
likelihood value.
This will be especially valuable in assessing the evidence for
stochasticity, where we will compare a model of perfect
correlation with one where $r_\LN \ne 1$ is allowed, effectively
introducing an extra parameter.
The key question is
how large a boost in likelihood is expected from the
introduction of an extra
parameter, and this was considered by \citet{astro-ph/0401198}. He
advocates the {\it Bayesian Information Criterion\/}, defined as
\be
B = -2\ln L + p \ln N,
\ee
where $p$ is the number of parameters and $N$ is the number of
data points. This measure of information effectively says that
going from a satisfactory model with $p$ parameters to one
that over-fits with $p+1$ parameters would be expected to increase
$\ln L$ by $0.5\ln N$. Therefore, in order to achieve evidence in
favour of the increase to $p+1$ at the usual 5~per~cent threshold,
we require 
\be\label{eq_liddle}
\Delta \ln L = -\ln 0.05 + 0.5\ln N,
\ee
which is between 5 and 8 for the number of cells considered
here. An unequivocal detection of stochasticity thus apparently requires
a likelihood ratio between $r_\LN \ne 1$ and the best $r_\LN =1$ model in
excess of $\lambda \simeq \exp(5)$ to $\exp(8)$.

Monte Carlo simulations may be used to check the validity
of this analytic method.
This is computationally expensive, so only an upper limit may be set
on the significance of an observed likelihood ratio. 
We create 40 mock datasets following a power law bias model
convolved with a Poisson distribution, defining the mean
cell counts, number of cells, one point PDF fit parameter $\sigma_\Ej$ and
model parameter $b_{\rm pow}$ to emulate a range of datasets.
To these we fit both power law and bivariate lognormal models with the
usual maximum likelihood fitting procedure. This allows us to 
assess the largest likelihood ratio that should arise by chance
if the true model is in fact perfect power law bias.
The results suggest a substantially smaller critical
value is required than equation (\ref{eq_liddle}), closer to $\Delta \ln
L = 1$ to reject the model at the $95\%$ confidence limit.
It therefore appears that the assumptions used to derive the 
Bayesian Information Criterion do not apply to this problem.

\subsubsection{Kolmogorov-Smirnov test}\label{sec_ks}
Although the likelihood ratio
test can eliminate one model in favour of another, it can not tell us
how well the preferred model fits the data.
A Kolmogorov-Smirnov (KS) test can be used to test for a difference
between an observed and modelled cumulative probability distribution. This test
provides the probability that the data are drawn from the model
probability distribution, with a low probability representing a poor
fit. A resulting probability above about 0.1 is generally
accepted as a reasonable fit, as the KS test is unable to rule
out the model being the true underlying distribution at greater than
90\% confidence. 
Strictly the test becomes
invalid once the data has been used to fix any free parameters of the
model, as in this method \citep{1993stp..book.....L}. However, as long
as the number of data points is much greater than the number of free
parameters any effects should be small.

To transform our bivariate distribution to a 1D variable on which we
can perform the standard KS test, we create integrated probability
distributions from both the model and the data, integrating within
constant model probability contours centered on the position of maximum probability. This gives
cumulative probability distributions for model and data from which the KS probability (that
the data follow the same underlying distribution as the model) can
be derived. 

The KS test has been generalised to bivariate analyses by
\citet{1983MNRAS.202..615P} and \citet{1987MNRAS.225..155F}. However,
this 2D KS test was found to lack power compared to the previous
method for the present application.

\section{Results}
\begin{table}
\centering
\caption{\small \label{tab1} Completeness limits, total number of cells and average
cell counts for each dataset after corrections for completeness have
been applied. Note that numbers do not scale exactly as $L^3$ due to
edge effects.}
\vspace{0.2cm}
\begin{tabular}[b]{ccccccc} \hline \hline
& Cell size (Mpc) & compl. & no. cells & $\langle N_\Ej\rangle$ &
$\langle N_\Lj\rangle$ \\ 
\hline
Colour 
& 10 & 0.7 & 11423& 1.5& 1.8  \\
& 15 & 0.7 & 3019& 5.1&6.3 \\
& 20 & 0.7 &  1104&12.1 & 14.6\\
& 25 & 0.7 & 484& 25.6 &30.1 \\
& 30  & 0.7 & 234 & 41.3& 48.6 \\
& 35  & 0.6 & 169& 57.4& 70.7 \\
& 40  & 0.6 & 115 & 88.3 & 105.4 \\
& 45 & 0.6 & 72 & 125.9& 149.2 \\ \hline
$\eta$  
& 10 & 0.7 & 9668& 1.9& 1.9\\
& 15 & 0.7 & 2567& 6.5&6.3\\
& 20 & 0.7 & 930& 15.3&14.7\\
& 25 & 0.7 & 404 & 32.1 & 30.2\\
& 30   & 0.7 & 187 & 54.2&49.6 \\
& 35   & 0.6 &  115 & 71.8 & 71.4 \\
& 40  & 0.6 &  74& 106.4& 108.7 \\
\end{tabular}

\end{table}

\begin{figure*}
\vspace*{-0.4cm}
\begin{minipage}{\textwidth}
\hspace*{-2.5cm}
\includegraphics[scale=0.9]{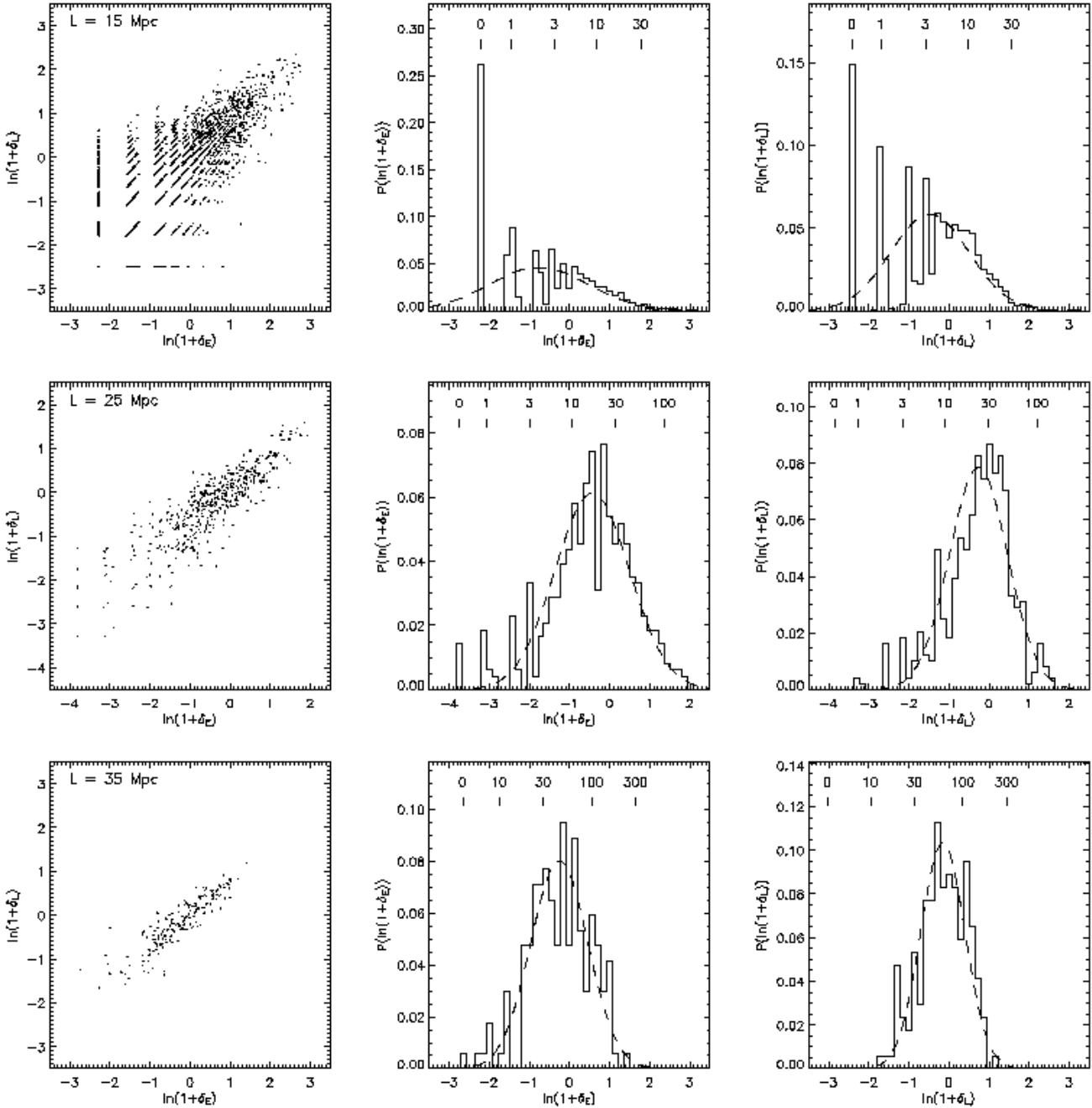}
\vspace*{-1.75cm}
\caption{\small On the left, the bivariate counts-in-cells
  distributions with early and late type galaxies classified by colour. 
  The points mark density values of individual cells, and from top to
  bottom $L=15$, 25 and $35\Mpc$ cells are shown. 
  1D projections of the distributions are shown for early types
  (centre) and late types (right), to
  which a Poisson sampled lognormal model has been fitted. The
  best--fitting lognormal curves are overplotted (dashed line). Due to
  the logarithmic axes, a bin for cells containing zero galaxies has been
  artificially positioned on the horizontal axis. Note the
  discreteness of the galaxy counts: the actual number of galaxies
  contained in the cells is indicated by the numbers over the 1D
  distributions. Further note the survey completeness 
  effects on smaller counts per cell, causing
  the spread of points around the mean value. Correcting zero counts
  for completeness is non-trivial and not included in this analysis,
  hence there is no spread of these points.
 }
\label{fig_1d}
\end{minipage}
\vspace*{-0.2cm}
\end{figure*}

Table \ref{tab1} summarises the details of each of our samples,
including average count per cell, and completeness limit.
The following Sections look in detail at the univariate and bivariate
model fits to each dataset. In Section \ref{sec_scale} we investigate
the scale dependence of nonlinearity and stochasticity in the 2dFGRS. In
Sections \ref{sec_origin} and \ref{sec_MC} we discuss
the origin of the stochasticity and perform some consistency checks
on the results.

\subsection{One point probability distributions}
\begin{figure*}
\vspace*{-0.5cm}
\begin{minipage}{\textwidth}
\hspace*{-0.5cm}
\includegraphics[scale=0.9]{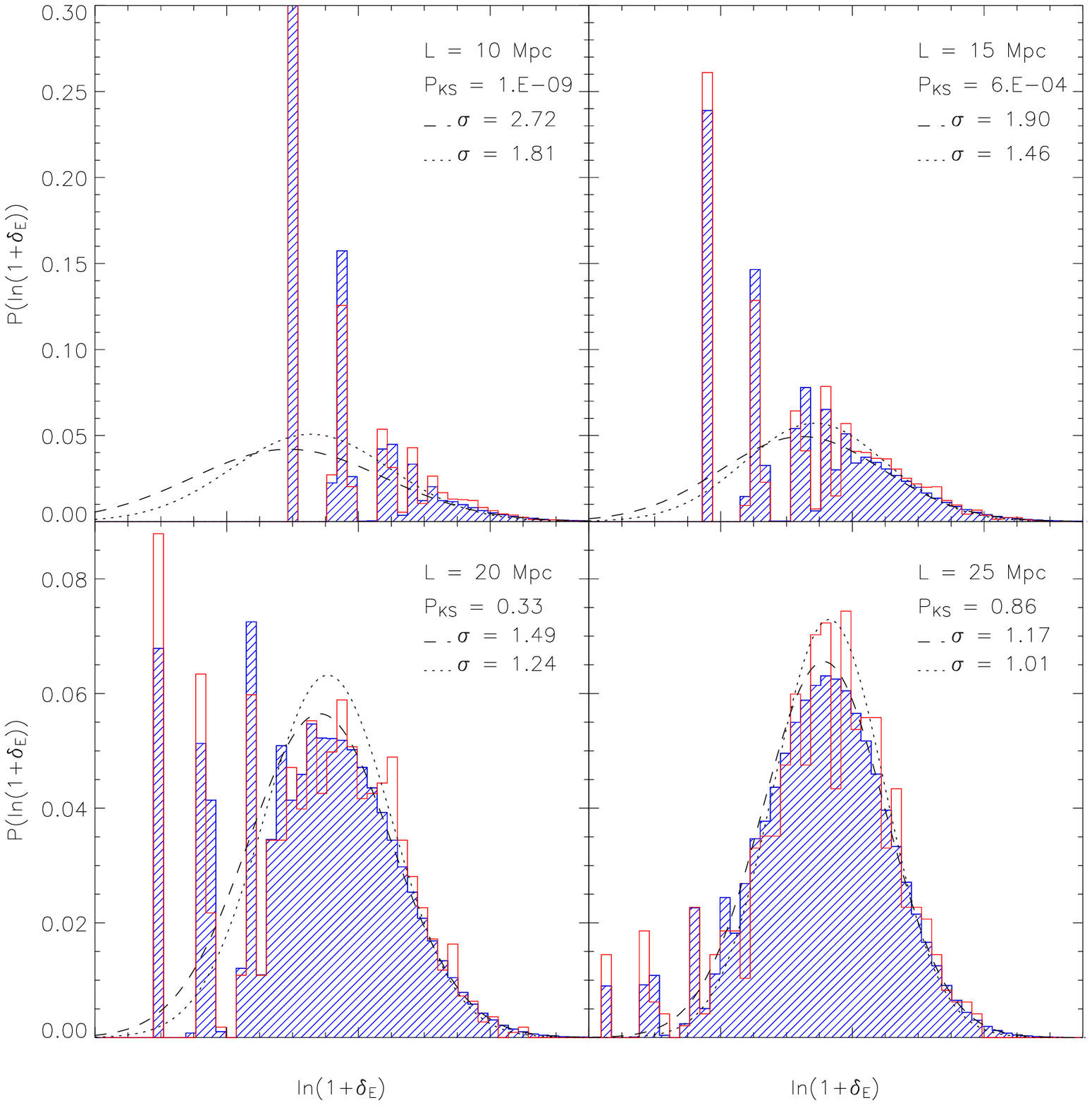}
\vspace*{-0.2cm}
\caption{\small The univariate distributions of early type galaxies
  for $L=10,15,20$ and $25\Mpc$ cells (empty, red histogram),
  together with the distribution of Monte Carlo cells (hatched, blue histogram) with parameters
  equal to those obtained from fitting a Poisson sampled lognormal
  distribution to the data cells. Completeness effects are modelled as
  a Gaussian with variance equal to that found in the dataset. The dashed
  curve shows this best-fitting 
  lognormal distribution; the dotted curve shows the lognormal curve
  with variance equal to that measured directly from the data. Both
  variances are given in the upper right, together with KS
  probabilities that the Monte Carlo cells are drawn from the same
  distribution as the data cells. Due to the logarithmic axes, a bin
  for cells containing zero galaxies has been artificially placed
  on the horizontal axis. For
  $L=10\Mpc$ about 60\% of the cells contain zero early type galaxies, and
  the vertical axis has been truncated to allow a better view of the
  remaining bins. Note the discreteness of the data and simulations,
  as discussed in the caption to Fig. \ref{fig_1d}. 
 }
\label{fig_1dMC}
\end{minipage}
\vspace*{-0.2cm}
\end{figure*}

Fig. \ref{fig_1d} shows the bivariate distributions of cell counts,
together with the one point distribution functions for a range of cell
sizes. Before we consider the bivariate distributions further, we look
in detail at the individual lognormal fits to these one point
distributions. We fit a lognormal distribution convolved with Poisson
noise to the early and late number counts individually, using the method
described in the previous Sections. The best fitting lognormal models
are shown overplotted in the Figure. It can be seen that on large
scales the lognormal model alone fits the data well, but on
small scales the deviation due to discreteness is substantial. For this
reason it is important to account for shot noise in the fitting
procedure. 

In order to assess the Poisson sampled lognormal model
quantitatively, we create many Monte Carlo cells with best fitting
parameters and expected number counts to match each dataset. We allow
for completeness effects by randomly assigning each cell a
completeness value from a Gaussian distribution of width and mean
equal to those of the dataset.  Fig. \ref{fig_1dMC} shows the
distributions of early type galaxies and Monte Carlo cells with
matching parameters. Overplotted are the best fitting lognormal model
(dashed line) and a lognormal curve with variance derived directly
from the data (dotted line, see Section \ref{efst90}). 

We can now compare our MC data with our true data through a KS
test. For large cells ($\ge 25\Mpc$) we find KS 
probabilities in excess of 0.8, but as cell size decreases the KS
probabilities decrease. On the smallest scales of $10\Mpc$ we obtain
KS probabilities of $\sim10^{-9}$ and it is this poor model fit that
causes the lognormal model to overestimate variances in comparison
with direct methods. The Figure shows there to be an excess of data
cells with moderate overdensities compared to the best-fitting
lognormal model, particularly on the smallest scales. In very
underdense regions the Figure shows the dotted (direct variance) curve
to lie below the dashed (fitted) curve. This results in an
underprediction of the number of cells containing zero or one
galaxy when using the direct variance method (as discussed in more
detail by Conway et al. 2004)

\subsubsection{Failures of the Poisson sampled lognormal distribution}\label{sec_1dfailure}
The discrepancy between the observed
and predicted distributions of cell counts shows that at least one of our
two assumptions about the galaxy field is incorrect. The lognormal
distribution is simply a convenient functional form which has been shown
to fit galaxy distributions from previous surveys
\citep[e.g.][]{1985ApJ...292L..35H,1994ApJ...420...44K} and N-body matter
distributions successfully \citep[][and references
therein]{1994ApJ...420...44K,2001ApJ...561...22K}.  
Deviations from this simple model are evident in
detailed numerical simulations \citep[e.g.][]{1995ApJ...443..479B}, and at
some level at least we would expect to see such deviations in our
data. Various alternative distributions have been suggested in the
literature such as the skewed lognormal, negative binomial or Edgeworth
expansions \citep[see also][]{1994ApJ...427..562S, astro-ph/0403593}. 

However, the fact that the model fails in detail on
scales at which shot noise dominates the distribution in underdense
regions suggests that the {\it Poisson sampling hypothesis\/} is at
least partly to blame. By attempting to fit the model to these
underdense cells, the variance is increased and the moderately
overdense regions are no longer well fitted. On small scales the
majority of cells contain zero or one galaxy, hence the preference of
the model to fit these cells and not those containing more galaxies.

We hope to explore more complex models for the count distribution
elsewhere. For the present application, there are two points to
make. The first is that cells of side about 10~Mpc are the smallest
that can sensibly be discussed with this approach; reducing the cell
size would lead to distributions that are dominated by discreteness
effects. More importantly, it should be stressed that analytic
models of this sort are not really physical. In the end, what
matters is whether the 2dFGRS data match the predictions of a
proper calculation of galaxy formation. We carry out such a comparison
at the end of the paper, and the results of a Poisson-sampled
lognormal fit are a convenient statistic to use for this purpose.
Provided true data and mock data are treated identically, small
imprecisions in the function used for the fit are irrelevant.

\subsection{Joint distributions and biasing models}

\begin{figure*}
\vspace*{-0.5cm}
\begin{minipage}{\textwidth}
\hspace*{-2cm}
\includegraphics[scale=0.85]{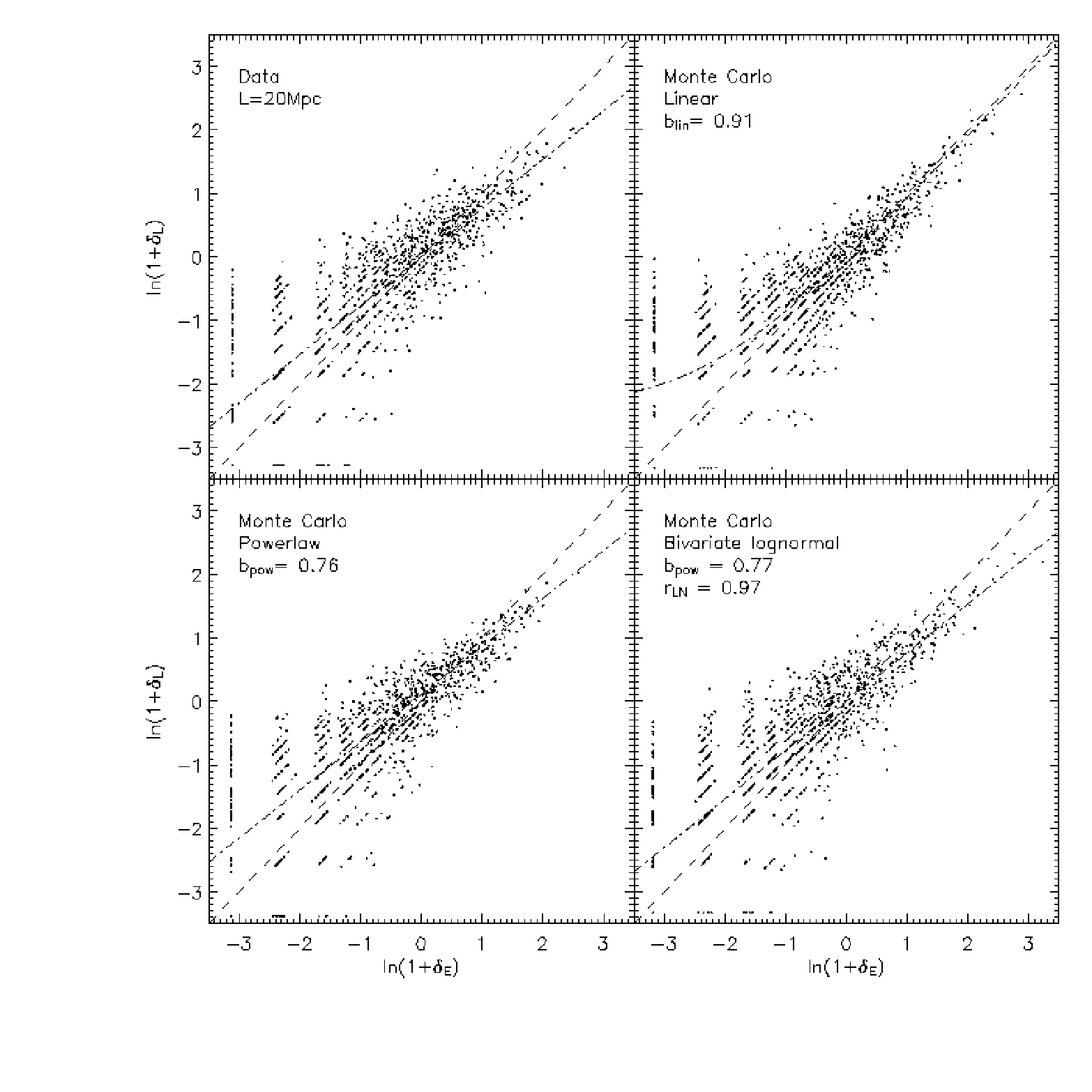}
\vspace*{-1.7cm}
\caption{\small On the top left, the bivariate counts-in-cells distributions for 
  $20\Mpc$ length cells, with early and late type galaxies classified by colour. 
  The points mark density values of individual cells. The other three panels
  show Monte Carlo realisations of the best fitting linear, power law and
  bivariate lognormal models. The realisations are created to match the
  data as far as possible, with equal cell numbers and average number
  counts. Cell completeness is included by assuming the distribution of
  cell completeness to be a Gaussian of mean and width equal to that of
  the data cells. In each panel the dashed line shows the $b=1.0$ case,
  and the dash-dot line shows the mean biasing of each
  model (for the top left plot, the dash-dot line shows the mean biasing of the
  best fitting bivariate lognormal model).
  Poisson sampling of the galaxies is assumed in all cases. 
  Note that for all but $b=1.0$, linear bias appears as a curve on the
  log-log plots. Due to the logarithmic axes, cells containing zero
  early or late type galaxies have been artificially positioned.}
\label{fig_col20}
\end{minipage}
\end{figure*}

\begin{figure*}
\vspace*{-0.5cm}
\begin{minipage}{\textwidth}
\hspace*{-2cm}
\includegraphics[scale=0.85]{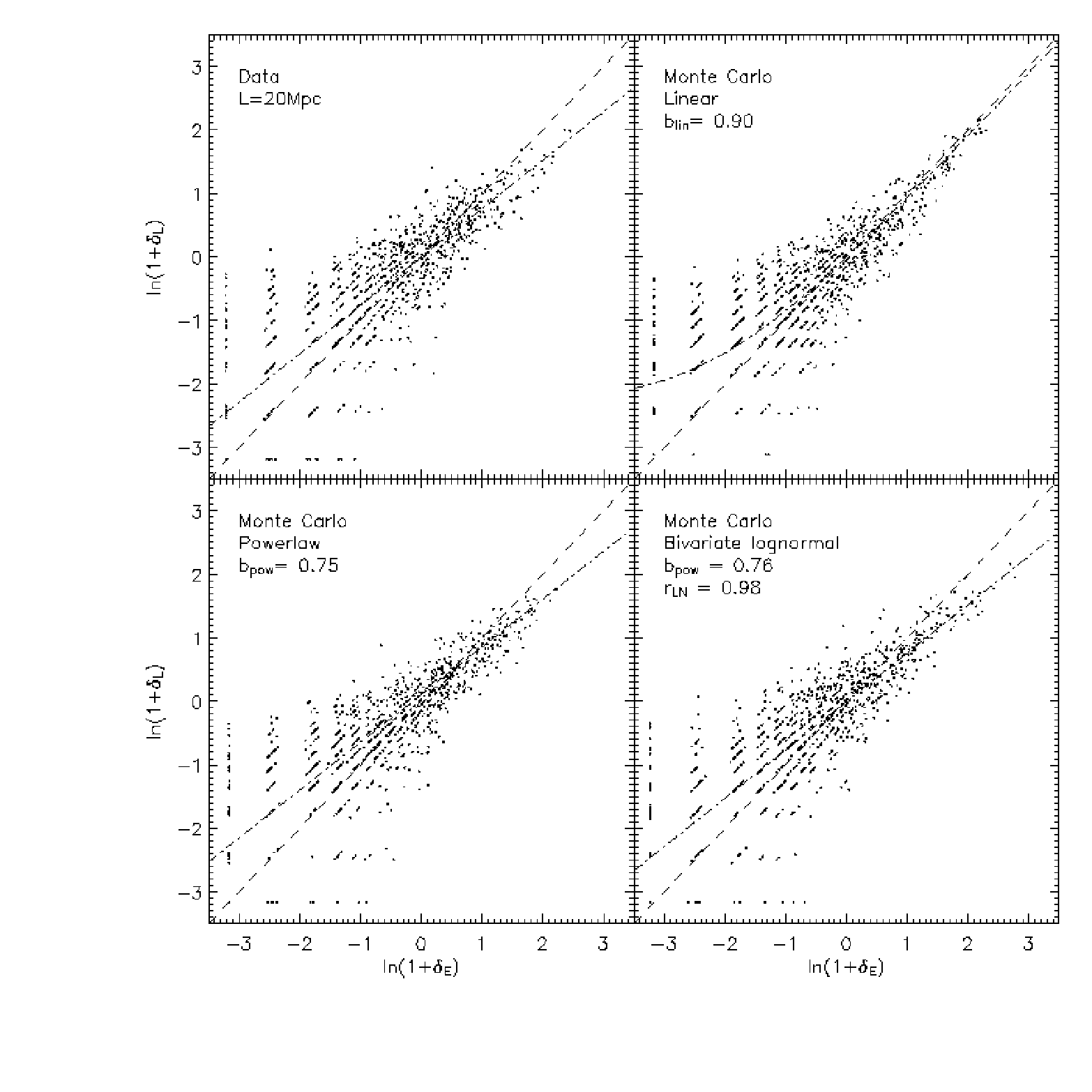}
\vspace*{-1.7cm}
\caption{\small Same as Fig. \ref{fig_col20}, with galaxies
  classified by $\eta$ type. }
\label{fig_eta20}
\end{minipage}
\end{figure*}

Each of the three models in Section \ref{models} is fitted
to the datasets described in Table \ref{tab1}. Best fit parameters
for the models are estimated simultaneously through the maximum likelihood method of
Section \ref{sec_like}.  Table \ref{tab2} shows the best fitting
parameter values for the two deterministic models, together with
log-likelihood differences between the model and the bivariate
lognormal model. 
The values of $b_{\rm pow}$ and
$b_{\rm lin}$ clearly show how early type galaxies are
more clustered than late type galaxies, as is well known. 
Table \ref{tab3} gives the best fitting parameter
values and errors for the stochastic bias model. The quoted errors are determined
through multivariate Gaussian fits to the likelihood surface, which
were found to agree well with Monte Carlo error estimates.

Fig. \ref{fig_col20} shows the joint probability distribution  of the
data for $L=20\Mpc$ cells, together with Monte Carlo realisations of the best fitting linear,
power law and bivariate lognormal models for comparison. The Monte Carlo realisations
include completeness effects by randomly selecting a completeness
value for each cell from a Gaussian with mean and width equal to that
of the true distribution of cell completeness. This scale is chosen to
illustrate all the properties of the data and the figures clearly show
the shot noise in underdense regions, together with the effects of
survey incompleteness. By eye we can see the differences between the
linear and power law bias models, and the effect of stochasticity in
the bivariate lognormal model. The best fitting linear model has a mean bias closer to
unity at high density, and cannot fit the nonlinearity seen in the data. The power law
model corrects for this, but the scatter about the mean is
insufficient to match the data. The stochasticity
introduced by the bivariate lognormal model is evident and is matched
well by the data.
The likelihood ratios shown in Table \ref{tab2} quantify the
differences and
show that on all scales the bivariate lognormal model gives
significantly better fits compared to deterministic biasing models.

We now repeat this analysis, splitting galaxies by spectral type $\eta$,
rather than by colour. The colour split 
allows a larger sample of cells to be included in the
analysis, but Fig.~\ref{fig_jointetacol} shows that
a division by spectral type does not always
select the same galaxies as a colour split, so it is interesting
to see how the results compare.
The second section of Tables \ref{tab1}, \ref{tab2} and
\ref{tab3} give details of the datasets and results of the model fits to cells with galaxies
classified by $\eta$. The joint distributions for $20\Mpc$ cells are
shown in Fig. \ref{fig_eta20}.

Comparing the results for $20\Mpc$ cells, the results for colour and
$\eta$ are generally similar. The
likelihood ratios again 
favour the bivariate lognormal model
over our two deterministic biasing models, and suggest a slightly
smaller difference between the stochastic and deterministic models
than found in the colour dataset. 
This is
verified by the smaller stochasticity found in the best fitting
bivariate lognormal model at all scales. Unlike for the colour datasets,
power law bias is only marginally inconsistent with the data on the largest scales
studied here. This difference between colour and $\eta$ type
results may reflect a physical difference in the relative biasing
relations, but firm conclusions are not yet possible.

\subsubsection{The goodness-of-fit statistics}
Table \ref{tab2} shows the log-likelihood differences $(-\ln\lambda)$ of the
parameter fits, taking the bivariate lognormal model as our null
hypothesis. In all cases the linear model provides a worse fit
to the data than the power law model, and the power law a worse fit than
the bivariate lognormal model. This latter statement may however be
due to the addition of an extra free parameter
to the model \citep{astro-ph/0401198}.
To establish the significance of the difference between the power law
and bivariate lognormal model we can make use of the theory given in Section
\ref{sec_modelcomp}.  For example, for $L=20\Mpc$ cells and assuming
the Bayesian Information Criterion
(equation \ref{eq_liddle}), we would require a likelihood
ratio in excess of $\exp(6.5)$ to claim that the bivariate lognormal model
provides a significantly better fit than the power law model with
$r_\LN=1$.  However, as stated earlier, Monte Carlo simulations of the power law model
show that a less stringent likelihood ratio in excess of $\simeq \exp(1)$ is all that
is required. 
We measure a likelihood ratio of $\exp(38)$ for the dataset
with galaxies classified by colour, and $\exp(18)$ for $\eta$
classification, a highly significant result in both cases.

Although the likelihood ratios favour of the bivariate lognormal
distribution over the two deterministic models, they do not tell us
how well the best-fitting distribution matches the data.
For this we turn to the KS statistic
described in Section \ref{sec_ks}.
On scales $L \ge 15\Mpc$ our KS statistic accepts the model with
a probability greater than 0.5. 
On the smallest scales studied here we find this probability to
decrease, in line with the trend for the univariate lognormal
distribution (Section 6.1).

\begin{table*}
\centering
\parbox{16cm}{\caption{\label{tab2} \small The best fitting
    deterministic biasing models parameters to each dataset.  The
    level of nonlinearity given the model is given by
    $\tilde{b}/\hat{b}$, which is unity by definition for the linear
    model. The penultimate column
    shows the log-likelihood differences between the best fit linear or power law
    models and bivariate lognormal model. A positive value indicates the
    bivariate lognormal model is a better fit to the data. The final
    column shows how many cells must be removed to reduce the power law
    likelihood ratio to $\sim \exp(1)$ (Section \ref{sec_origin}).}}
\vspace{0.2cm}
\begin{tabular}[htb]{ccccccccccc} \hline \hline
 & Cell size & Model & $\omega_\Ej$ &  $b_{\rm lin}$ or $b_{\rm pow}$
 & $\hat{b}$ & $b_{\rm var}$ & $r_{\rm lin}$ & $\tilde{b}/\hat{b}$ &
 $\mathcal{L}-\mathcal{L}^{\rm LN}$ & Outliers\\  \hline
Colour
& 10 & linear& 1.41 & 0.93 & 0.93 & 0.93  & 1.00 & 1.000 & 291.1 &\\
 && power law & 1.51 & 0.78 & 0.56 & 0.58 & 0.96 & 1.044 &   75.6 &159\\
& 15 & linear& 1.23 & 0.91 & 0.91 & 0.91  & 1.00 & 1.000 & 185.0 &\\
 && power law & 1.26 & 0.77 & 0.61 & 0.63 & 0.97 & 1.030 &   55.5 &86\\
& 20 & linear& 1.04 & 0.91 & 0.91 & 0.91  & 1.00 & 1.000 & 133.6 &\\
 && power law & 1.10 & 0.76 & 0.63 & 0.65 & 0.98 & 1.024 &   37.9 &40\\
& 25 & linear& 0.92 & 0.91 & 0.91 & 0.91  & 1.00 & 1.000 &  87.1 &\\
 && power law & 0.94 & 0.76 & 0.67 & 0.68 & 0.98 & 1.016 &   41.3 &34\\
& 30 & linear& 0.80 & 0.92 & 0.92 & 0.92  & 1.00 & 1.000 &  39.3 &\\
 && power law & 0.77 & 0.77 & 0.72 & 0.72 & 0.99 & 1.009 &   22.4 &18\\
& 35 & linear& 0.73 & 0.92 & 0.92 & 0.92  & 1.00 & 1.000 &  16.0 &\\
 && power law & 0.70 & 0.76 & 0.71 & 0.72 & 0.99 & 1.008 &    6.6 &6\\
& 40 & linear& 0.67 & 0.95 & 0.95 & 0.95  & 1.00 & 1.000 &  18.1 &\\
 && power law & 0.68 & 0.81 & 0.77 & 0.78 & 1.00 & 1.005 &   12.4 &8\\
& 45 & linear& 0.59 & 0.97 & 0.97 & 0.97  & 1.00 & 1.000 &  12.9 &\\
 && power law & 0.59 & 0.86 & 0.83 & 0.83 & 1.00 & 1.002 &   10.4 &1\\
\hline
$\eta$ 
& 10 & linear& 1.43 & 0.92 & 0.92 & 0.92  & 1.00 & 1.000 & 238.1 &\\
 && power law & 1.51 & 0.77 & 0.55 & 0.58 & 0.96 & 1.046 &   49.8 &110\\
& 15 & linear& 1.22 & 0.91 & 0.91 & 0.91  & 1.00 & 1.000 & 131.1 &\\
 && power law & 1.24 & 0.77 & 0.64 & 0.65 & 0.97 & 1.026 &   35.6 &58\\
& 20 & linear& 1.03 & 0.90 & 0.90 & 0.90  & 1.00 & 1.000 &  91.9 &\\
 && power law & 1.07 & 0.75 & 0.67 & 0.68 & 0.98 & 1.019 &   17.9 &25\\
& 25 & linear& 0.93 & 0.90 & 0.90 & 0.90  & 1.00 & 1.000 &  70.3 &\\
 && power law & 0.96 & 0.74 & 0.65 & 0.67 & 0.98 & 1.018 &   23.7 &21\\
& 30 & linear& 0.81 & 0.90 & 0.90 & 0.90  & 1.00 & 1.000 &  28.4 &\\
 && power law & 0.81 & 0.73 & 0.67 & 0.68 & 0.99 & 1.013 &   13.3 &9\\
& 35 & linear& 0.74 & 0.90 & 0.90 & 0.90  & 1.00 & 1.000 &  13.6 &\\
 && power law & 0.75 & 0.69 & 0.65 & 0.66 & 0.99 & 1.013 &    4.3 &2\\
& 40 & linear& 0.64 & 0.95 & 0.95 & 0.95  & 1.00 & 1.000 &   4.8 &\\
 && power law & 0.65 & 0.83 & 0.82 & 0.82 & 1.00 & 1.003 &    1.2 &0\\
\end{tabular}
\end{table*}

\begin{table*}
\centering
\parbox{15.75cm}{\caption{\label{tab3} \small The best-fitting 
    bivariate lognormal model parameters to each dataset. Errors are
    shown, derived from Gaussian fits to the parameter likelihood
    surface. $\Delta(r_{\LN})$ is derived from propagation of
    $\smash{\Delta[(1-r_{\LN}^2)^{1/2}]}$. The remaining columns give the
    average biasing parameters. Appendix \ref{app1} gives the analytic
    solutions for each parameter in the case of the bivariate lognormal model. The
    final two parameters measure the nonlinearity
    and stochasticity of the model (equations \ref{eq_nonlin},
    \ref{eq_stoc}).}}
\vspace{0.2cm}
\begin{tabular}[htb]{ccccccccccccccc} \hline \hline
& Cell size & $\omega_\Ej$ & $\Delta(\omega_\Ej)$ &
  $\omega_\Lj$ & $\Delta(\omega_\Lj)$ & $r_{\LN}$ & $\Delta(r_{\LN})$ &
  $\sigma_\Ej$ & $\sigma_\Lj$  & $r_{\rm lin}$ & $\hat{b}$ &
  $b_{\rm var}$ & $\tilde{b}/\hat{b}$ &
  $\sigma_b/\hat{b}$ \\ \hline
Colour 
&10 & 1.52 &  0.01 & 1.20 &  0.01 & 0.958 & 0.004 & 3.01 & 1.80 & 0.88 & 0.52 & 0.55 & 1.054 & 0.44 \\
&15 & 1.26 &  0.02 & 0.99 &  0.02 & 0.966 & 0.004 & 1.99 & 1.29 & 0.92 & 0.60 & 0.62 & 1.033 & 0.35 \\
&20 & 1.10 &  0.02 & 0.85 &  0.02 & 0.969 & 0.005 & 1.54 & 1.02 & 0.93 & 0.62 & 0.64 & 1.026 & 0.31 \\
&25 & 0.95 &  0.02 & 0.73 &  0.02 & 0.959 & 0.007 & 1.21 & 0.84 & 0.93 & 0.64 & 0.66 & 1.020 & 0.34 \\
&30 & 0.78 &  0.03 & 0.61 &  0.03 & 0.962 & 0.009 & 0.92 & 0.68 & 0.94 & 0.69 & 0.70 & 1.011 & 0.31 \\
&35 & 0.71 &  0.04 & 0.54 &  0.03 & 0.976 & 0.009 & 0.81 & 0.58 & 0.96 & 0.70 & 0.70 & 1.009 & 0.24 \\
&40 & 0.68 &  0.05 & 0.55 &  0.04 & 0.971 & 0.009 & 0.76 & 0.60 & 0.96 & 0.75 & 0.75 & 1.006 & 0.27 \\
&45 & 0.60 &  0.04 & 0.51 &  0.04 & 0.970 & 0.010 & 0.66 & 0.55 & 0.96 & 0.80 & 0.80 & 1.003 & 0.27 \\
\hline
$\eta$
&10 & 1.51 &  0.02 & 1.18 &  0.01 & 0.963 & 0.005 & 2.95 & 1.75 & 0.89 & 0.53 & 0.56 & 1.052 & 0.40 \\
&15 & 1.23 &  0.02 & 0.98 &  0.02 & 0.966 & 0.005 & 1.89 & 1.27 & 0.92 & 0.62 & 0.64 & 1.029 & 0.34 \\
&20 & 1.07 &  0.02 & 0.82 &  0.02 & 0.976 & 0.005 & 1.47 & 0.98 & 0.94 & 0.63 & 0.64 & 1.025 & 0.27 \\
&25 & 0.92 &  0.02 & 0.70 &  0.02 & 0.966 & 0.007 & 1.16 & 0.79 & 0.94 & 0.64 & 0.65 & 1.019 & 0.31 \\
&30 & 0.81 &  0.05 & 0.60 &  0.04 & 0.965 & 0.010 & 0.97 & 0.66 & 0.94 & 0.64 & 0.65 & 1.016 & 0.30 \\
&35 & 0.73 &  0.04 & 0.51 &  0.03 & 0.980 & 0.009 & 0.84 & 0.55 & 0.96 & 0.63 & 0.64 & 1.015 & 0.22 \\
&40 & 0.66 &  0.04 & 0.54 &  0.04 & 0.988 & 0.008 & 0.73 & 0.58 & 0.98 & 0.78 & 0.79 & 1.004 & 0.17 \\

\end{tabular}
\end{table*}

\begin{figure*}
\vspace*{-0.5cm}
\begin{minipage}{\textwidth}
\hspace*{-2cm}
\includegraphics[scale=0.9]{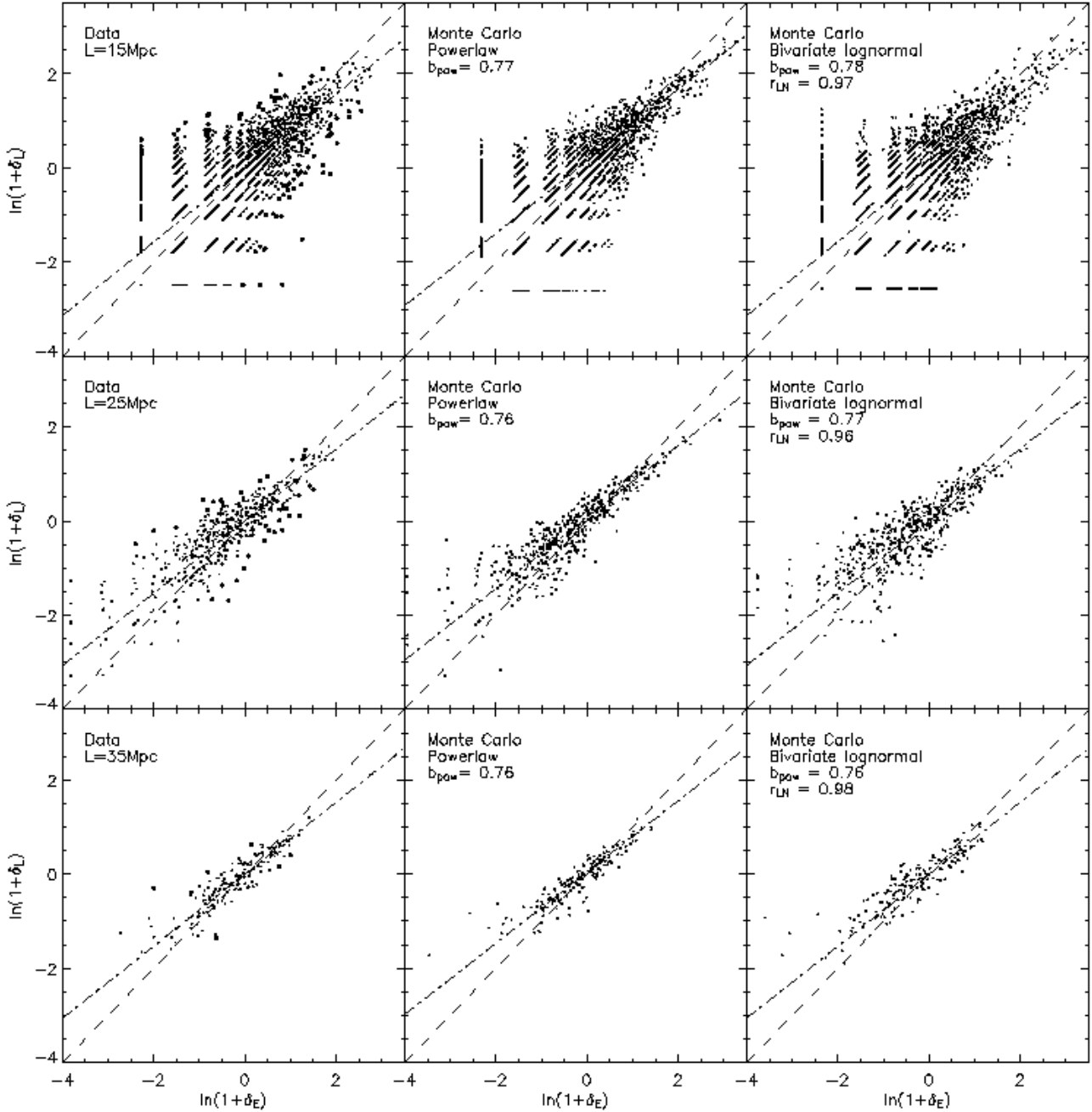}
\vspace*{-2.2cm}
\caption{\small The bivariate counts-in-cells distributions for 
  15 (top), 25 and $35\Mpc$ length cells. The left
  hand panel shows the data with early and late galaxies classified by
  colour. Larger points indicate the cells identified as outliers from
  $r_\LN = 1$ (see Section \ref{sec_origin}). The central column shows a Monte Carlo simulation of the
  best fitting power law model and the right hand column the best fitting
  bivariate lognormal model. The dashed line indicates a mean biasing of
  $b=1.0$, the dot-dash line shows the best fit mean bias.}
\label{fig_colsd}
\end{minipage}
\vspace*{-0.1cm}
\end{figure*}

\subsubsection{Stochasticity and nonlinearity}

\begin{figure*}
  \vspace*{-0.5cm}
\begin{minipage}{\textwidth}
\centering
\includegraphics[scale=0.66]{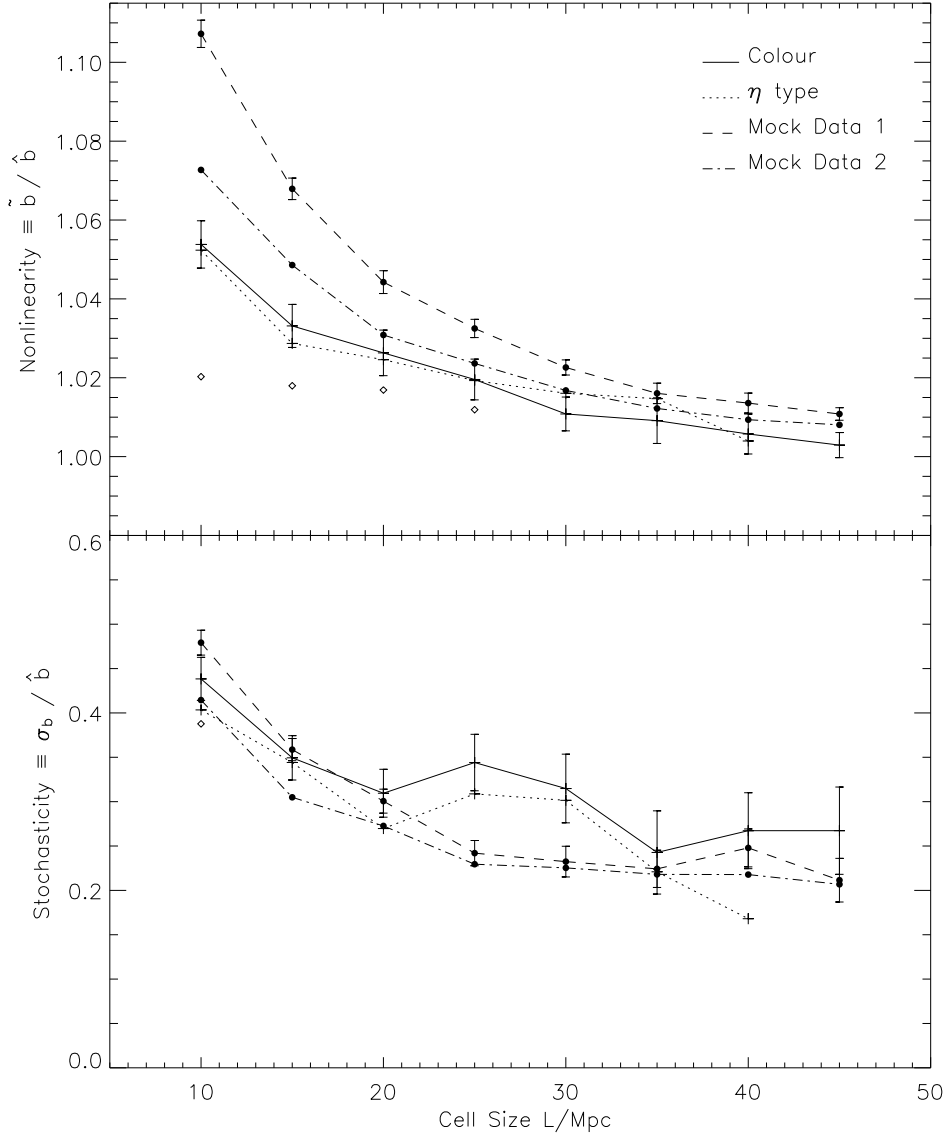}
\vspace*{-0.1cm}
\caption{\small The scale dependence of nonlinearity and stochasticity
in the 2dFGRS. The solid line shows results for galaxies classified by
colour and dotted line for galaxies classified by $\eta$ type. The circles show
the two semianalytic datasets. Mock 1 is described in the text as the
``superwind'' model, and mock 2 as the ``low-baryon'' model. The open
diamonds indicate values measured for the colour dataset using direct
variance estimates, where they differ by more than 1 sigma from the results derived from
model fitting (see Section \ref{efst90}). For clarity, errors are omitted for the
$\eta$ and second mock datasets.}
\label{fig_nonstoc}
\end{minipage}
\end{figure*}

In order to quantify the nonlinearity and stochasticity of the joint
distribution of early and late type galaxies we assume that the bivariate
lognormal model is an accurate representation of the data. In our
analysis the log-density correlation coefficient $r_\LN$
provides a complete measure of the stochasticity; 
to aid comparison with other work we compute the 
mean biasing, its nonlinearity and the average biasing
scatter of equations (\ref{eq_hatb}) to (\ref{eq_sigb}). 
For clarity we concentrate briefly on the results for $20\Mpc$ cells, shown in
the third line of Table \ref{tab3}. These indicate that whilst the
nonlinearity [equation (\ref{eq_nonlin})] is only 1.03, the
stochasticity [equation (\ref{eq_stoc})] is 
0.31. This high stochasticity is reflected in the deviation of
$r_\LN$ from one, and the low linear correlation
coefficient of $r_{\rm lin}=0.93$. It is important to note that these
statistics account for Poisson noise, as the models were convolved
with a Poisson distribution before being fitted to the data.

Table \ref{tab2} shows for comparison some biasing statistics for our two
deterministic models. It can be seen that a similar nonlinearity is
measured by the power law model, whilst the linear correlation
coefficient remains close to one, reflecting the inability of the
model to measure stochasticity.  
The best fitting linear bias model has a mean biasing
parameter closer to one, indicating that by assuming this model previous studies may
have underestimated the magnitude of relative biasing. 

It may be considered surprising that the correlation
parameter $r_\LN$ can be measured so precisely that 
$r_\LN=0.97$ can be clearly distinguished from $r_\LN=1$.
The reason for this can be seen by examining the expression
for the bivariate lognormal distribution [equation (\ref{eq_bivLNcond})], in which the
scatter in $\delta_\Lj$ at fixed $\delta_\Ej$ is proportional
to $S \equiv \sqrt{1-r_\LN^2}$. 
This is a more meaningful quantity
than the correlation coefficient, but it lacks a standard name.
In this context, the obvious term for $S$
would be `stochasticity', but this is already taken and
we resist the temptation to expand the terminology further.
The stretched nature of this measure of correlation is
quite extreme (as noted independently by \citet{astro-ph/0403698}): $S=0.5$ corresponds to
$r_\LN=0.87$. Therefore, $r_\LN=0.87$ is effectively half-way
to no correlation at all. This is why even a correlation as
high as $r_\LN=0.97$ is noticeably imperfect in terms
of density-density plots.

\subsubsection{Scale dependence}\label{sec_scale}
\begin{figure}
\includegraphics[scale=0.4]{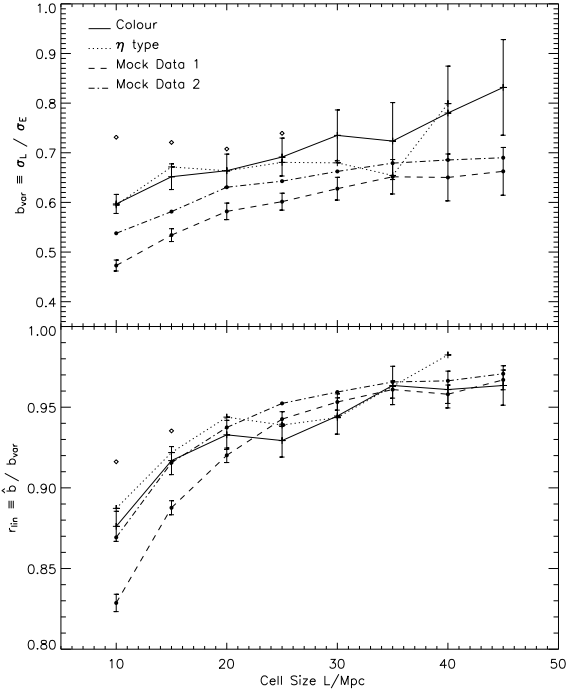}
\caption{\small The scale dependence of the ratio of variances,
  $b_{\rm var}$, and the linear
  correlation coefficient, $r_{\rm lin}$. Symbols as in Fig. \ref{fig_nonstoc}. }
\label{fig_bvarrlin}
\end{figure}

\begin{table*}
\centering
\parbox{9.25cm}{\caption{\label{tab5} \small Colour and $\eta$ samples with cell length
  $L=20\Mpc$ are each split into two
  redshift groups at $z = 0.09$. This Table shows results for the bivariate
  lognormal model fit to each subsample.}}
\vspace{0.2cm}
\begin{tabular}[htb]{cccccccc} \hline \hline
  &  & $\omega_\Ej$ & $\Delta(\omega_\Ej)$ &
  $\omega_\Lj$ & $\Delta(\omega_\Lj)$ & $r_\LN$ &  $\Delta(r_\LN)$ \\  \hline
Colour & low $z$ & 1.11 & 0.03 & 0.85 & 0.03 & 0.956 & 0.010 \\
       & high $z$ &1.11 & 0.03 & 0.85 & 0.03 & 0.974 & 0.006  \\
$\eta$ & low $z$ & 1.10 & 0.03 & 0.85 & 0.03 & 0.985 & 0.007 \\
& high $z$ &1.07 & 0.03 & 0.82 & 0.03 & 0.976 & 0.007 \\
\end{tabular}
\end{table*}

We now look at how our results depend on scale. This is interesting
because it may potentially distinguish whether the efficiency of
galaxy formation in a particular region of space is affected by local
or non-local factors.
Examples of local factors could be density, geometry,
or velocity dispersion of the dark matter. Non-local factors
could involve, for example, effects of ionising radiation from the
first stars or QSOs on galaxy formation efficiency, causing coherent
variation over larger scales than possible from local factors. 

Fig. \ref{fig_colsd} shows the bivariate distributions for 15, 25 and $35\Mpc$
cells with galaxies split by colour. The likelihood ratio tests (Table
\ref{tab2}) show that the
bivariate lognormal bias model provides a significantly better fit to
our data  than both deterministic models on all scales for colour
selection and all but the largest scales when classifying by
$\eta$. 

The nonlinearity and stochasticity as a function of scale are plotted
in Fig. \ref{fig_nonstoc}, with errors derived by propagation
from those shown in Table \ref{tab3}. The commonly quoted parameters
$b_{\rm var}$ and $r_{\rm lin}$, which combine both nonlinearity and
stochasticity, are plotted in Fig. \ref{fig_bvarrlin} to ease
reference with results in the literature. On small scales ($\le
20\Mpc$) the average biasing statistics suffer systematic errors from
overestimates of the variance by the Poisson sampled lognormal model fit, as discussed
in detail in Section \ref{efst90}. To indicate the magnitude of these
effects, open diamonds show results for
the colour dataset using direct variance estimates, 
where the difference is greater than 1 sigma. It can be seen that
although both nonlinearity and $b_{\rm var}$ show noticeable change with
scale, this can be mostly explained by the poor fit of the model. There is
little effect on stochasticity, and both mocks and data are affected
in the same way, making comparison practical.
The nonlinearity reaches $<1\%$ by around $35\Mpc$ with results for
colour and $\eta$ classification barely distinguishable. 
A little care is needed in interpreting this result, however:
negligible `nonlinearity' does not mean that linear bias
is a good fit. As much as anything, this is a statement that the amplitude
of fluctuations declines for large $L$, so most cells have $|\delta| \ll 1$.

The stochasticity also
declines, although on large scales the errors prevent distinction
between a flat or declining function with scale. 
There is a tendency for the stochasticity of the $\eta$
datasets to lie a little below that of the colour datasets, but
this is not significant within the errors. The dashed and dash-dot lines show the
results for two semianalytic mock universes which will be discussed in
detail in Section \ref{sec_dur}.
We can immediately note the encouraging general agreement: stochasticity
is clearly expected at about the detected level.

\subsubsection{Division by luminosity and redshift}\label{sec_zdep}

By splitting galaxies by their luminosity we can investigate whether
the effects that we find in the previous Sections could be due to the
luminosity difference of the galaxy types. By dividing galaxies at
$M-5\log_{10}(h)= -19.5$, we form two similar sized groups with class 1
being more luminous than class 2. The models are fitted as before by
replacing E (L) with class 1 (2). In contrast to the outcome when
galaxies are divided by type, the likelihood ratios between the best fitting
power law and bivariate lognormal models are small on all scales, ranging
from 0 to 3. We find $r_\LN$ to be roughly constant with a value of
$\sim 0.99$ for the best fitting bivariate lognormal models.  
If stochasticity is caused by some variable other than the local
density during galaxy formation, then perhaps luminosity is less dependent
on this variable than galaxy type and colour. Other explanations could
be that our volume limited sample is too shallow to find the expected
bimodality in luminosity, and the position of our boundary between bright
and faint galaxies is arbitrary.

\begin{figure*}
\begin{minipage}[t]{17cm}
\begin{minipage}{8.5cm}
\includegraphics[scale=0.65]{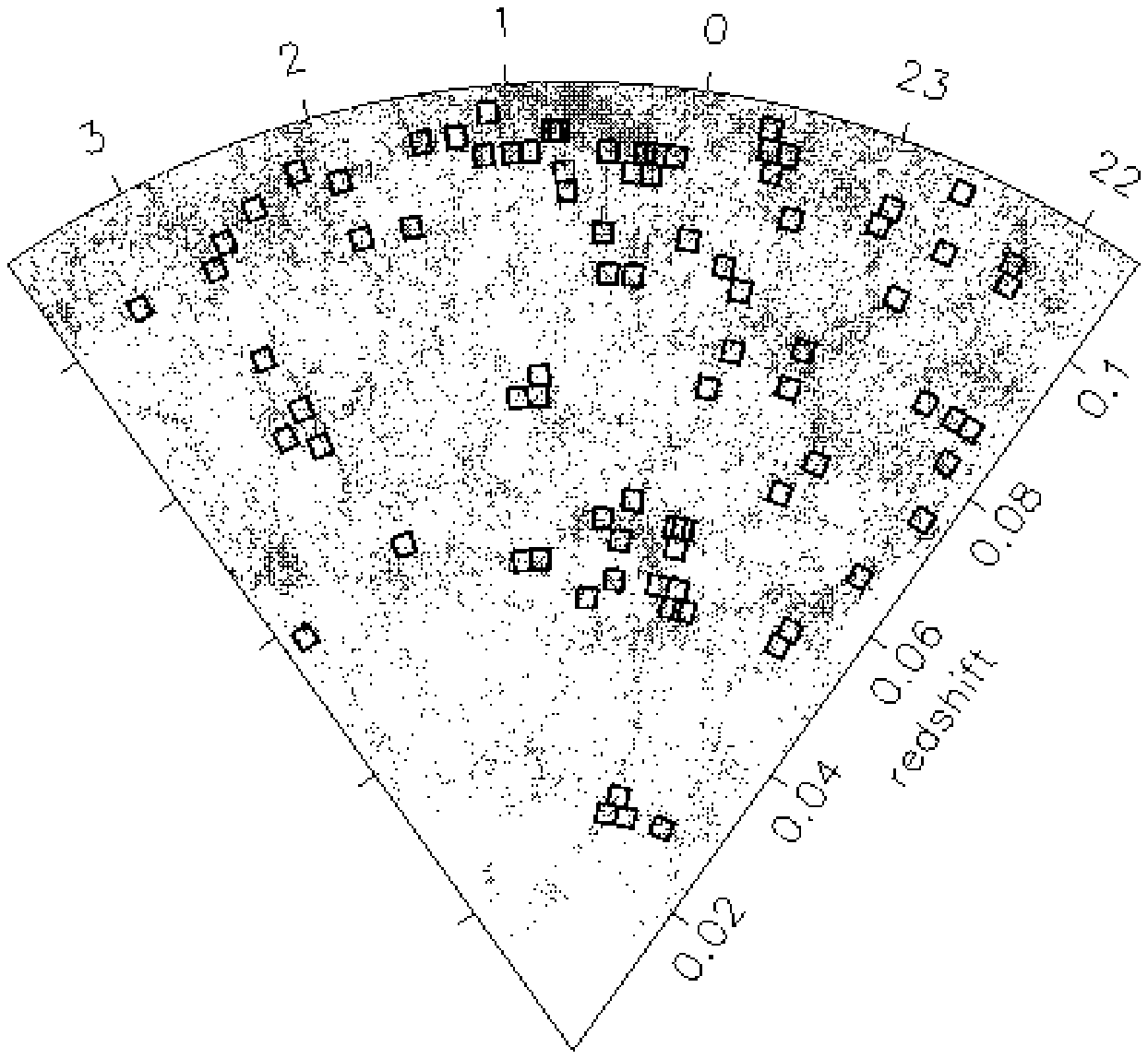}
\end{minipage}
\begin{minipage}{8.5cm}
\includegraphics[scale=0.65]{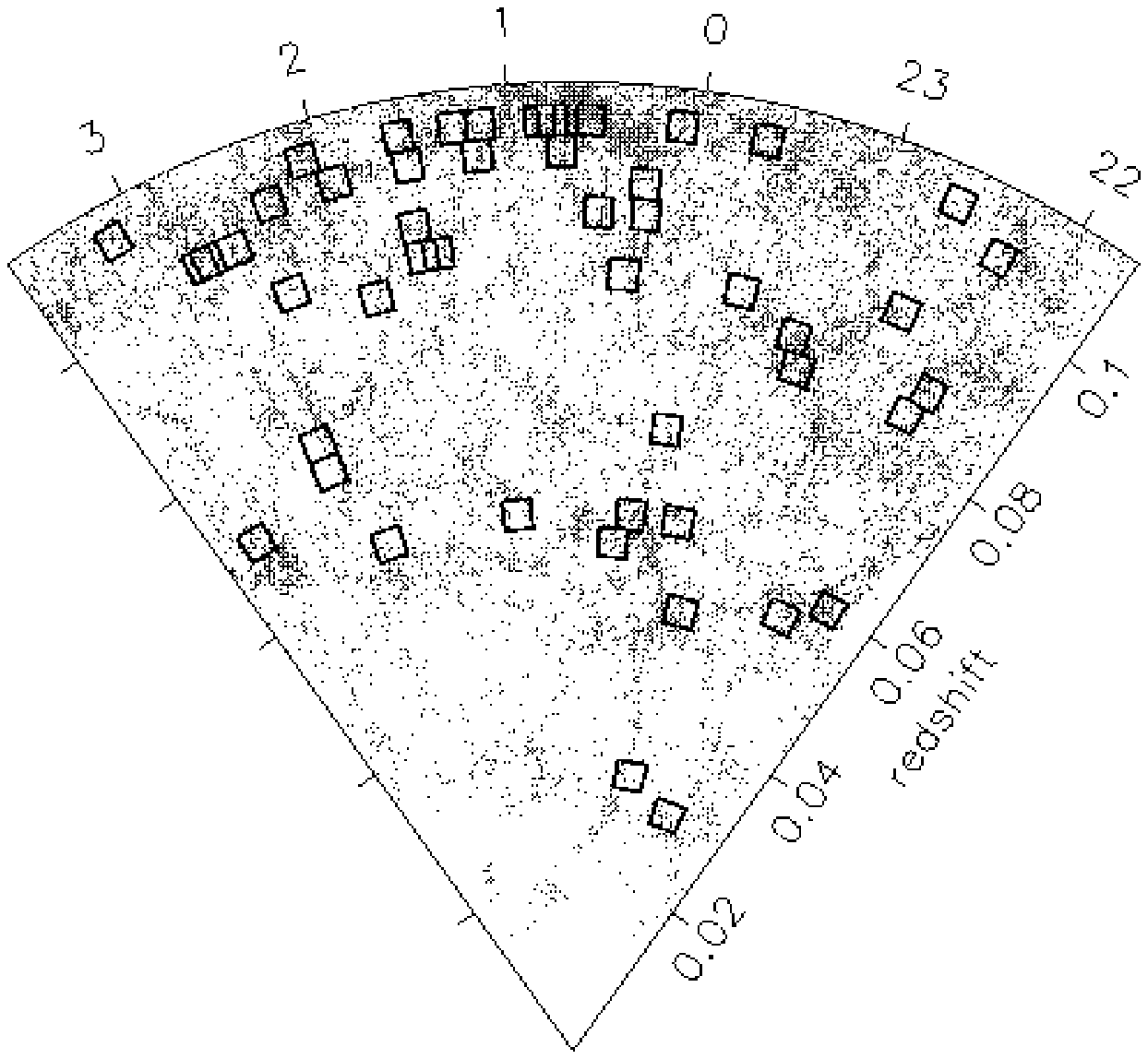}
\end{minipage}

\begin{minipage}{8.5cm}
\includegraphics[scale=0.65]{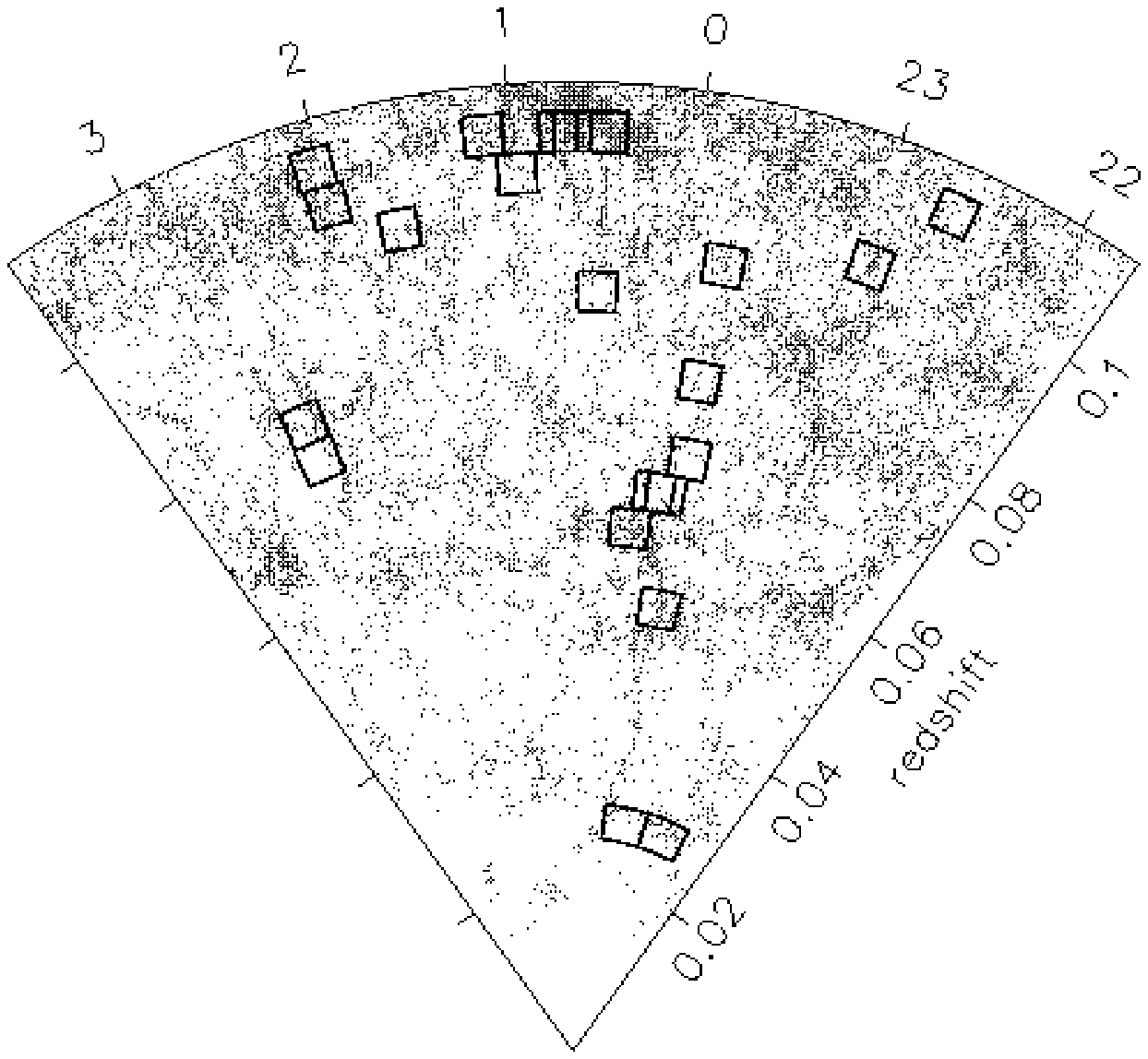}
\end{minipage}
\begin{minipage}{8.5cm}
\includegraphics[scale=0.65]{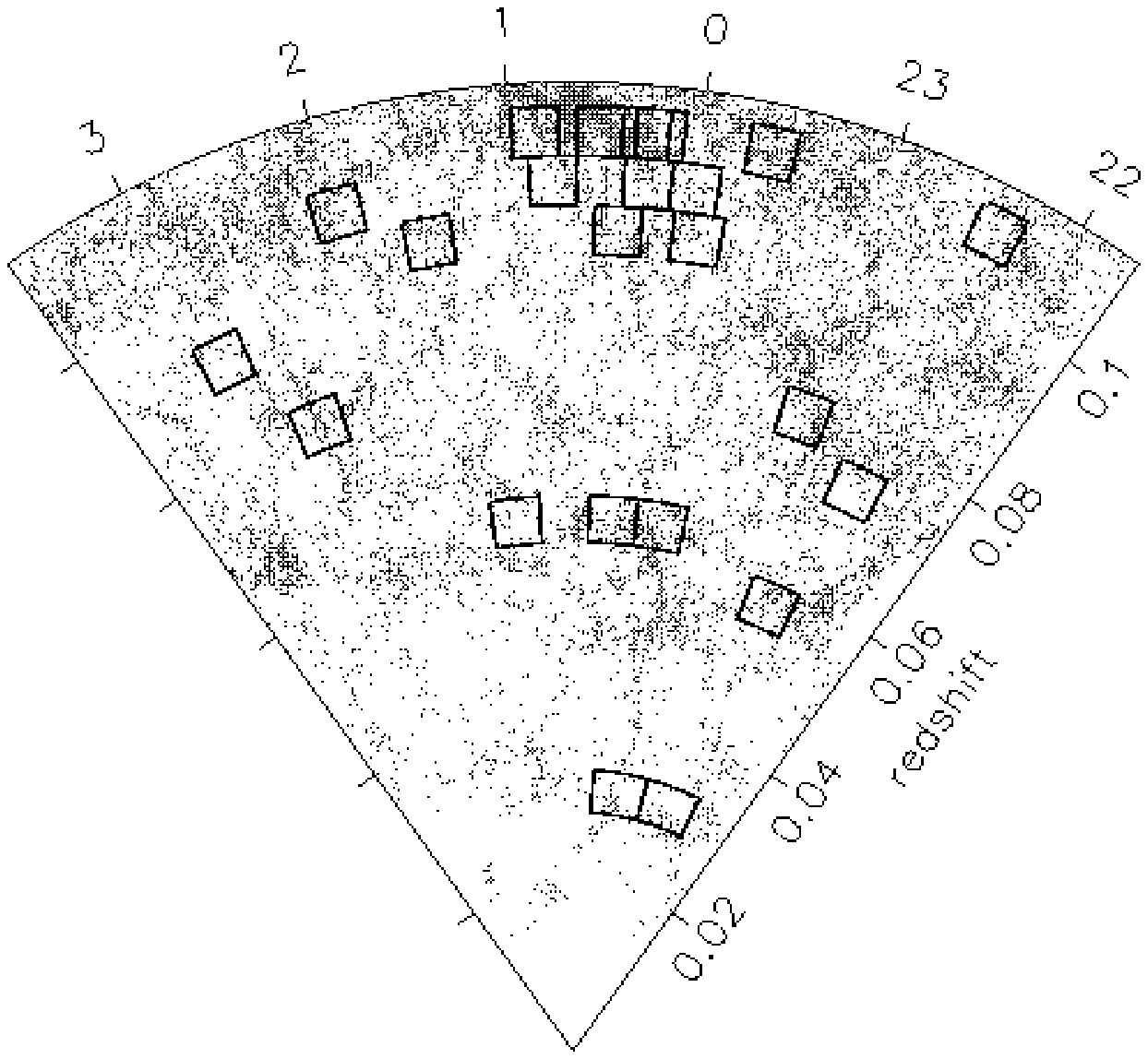}
\end{minipage}
\caption{\small Wedge plots of the 2dFGRS volume limited survey SGP
  region as for Fig.\ref{fig_boxes}. Both early and late type galaxies
  are shown. Overplotted are cells identified
  as causing the stochasticity signal, from top left:
  10,15,20,25$\Mpc$ cells.
  }
\label{fig_outliers}
\end{minipage}
\end{figure*}

It is also of interest to see if the results are independent of redshift.
We divide the survey at $z= 0.09$ and fit the models to both high and low
redshift galaxies using a cell length of $L = 20\Mpc$. Table
\ref{tab5} shows the best fitting bivariate lognormal parameters 
for galaxies split by colour and $\eta$ for both redshift groups.
Due to the fibre apertures of the 2dF instrument we may expect some
redshift dependence for galaxies classified by $\eta$ (see Section
\ref{sec_eta}), yet precise predictions are difficult. We certainly
see no difference within the errors between these two redshift groups,
and the difference for colour classification can
not be attributed to such effects. It is possible that the changing errors on
the colour at high redshift contribute to the decrease in
stochasticity, although evolution can not be ruled out. There is
certainly room for further investigation with forthcoming larger
redshift surveys.

\subsubsection{Comparison with other 2dFGRS results}
This work has been carried out in conjunction with that of Conway et
al. (2004) who investigate the variance and deviation from linear bias
in the 2dFGRS NGP and SGP regions using flux-limited samples,
including a counts-in-cells analysis. They find similar discrepancies
between the Poisson sampled lognormal model and the data,
investigating the causes and magnitude of the problem in detail. After
accounting for this effect in both analyses the results agree within
the 1 sigma errors where comparable: Conway et~al. find
$1/b_{\rm var}=1.25\pm0.05$, and nonlinearity ($\tilde{b}/\hat{b}$) of a few
percent on the smallest scales measured. Our results for $b_{\rm pow}$ agree,
but are consistently higher for $b_{\rm lin}$. This is due to
different fitting procedures; Conway et al. give greater weight to overdense
regions.

\citet{2003MNRAS.344..847M} measure the square root of the ratio of the correlation
functions of early and late type galaxies to be around 1.2 on their largest
scales of $8 < r < 20\mpcoh$. Their bias parameter corresponds
to $\smash{1/\sqrt{\hat{b}}}$ in the notation of Section
\ref{sec_frame} (DL99). This gives a value for $\hat{b}$ a little
higher than our
results of Table \ref{tab3}, but entirely consistent when lognormal
variance estimates are replaced by direct measures as in Section
\ref{sec_scale}. 

\subsection{Origin of the stochasticity signal}\label{sec_origin}

\begin{figure*}
\vspace*{-0.4cm}
\begin{minipage}[tb]{\textwidth}
\begin{minipage}{8.0cm}
\includegraphics[scale=0.6]{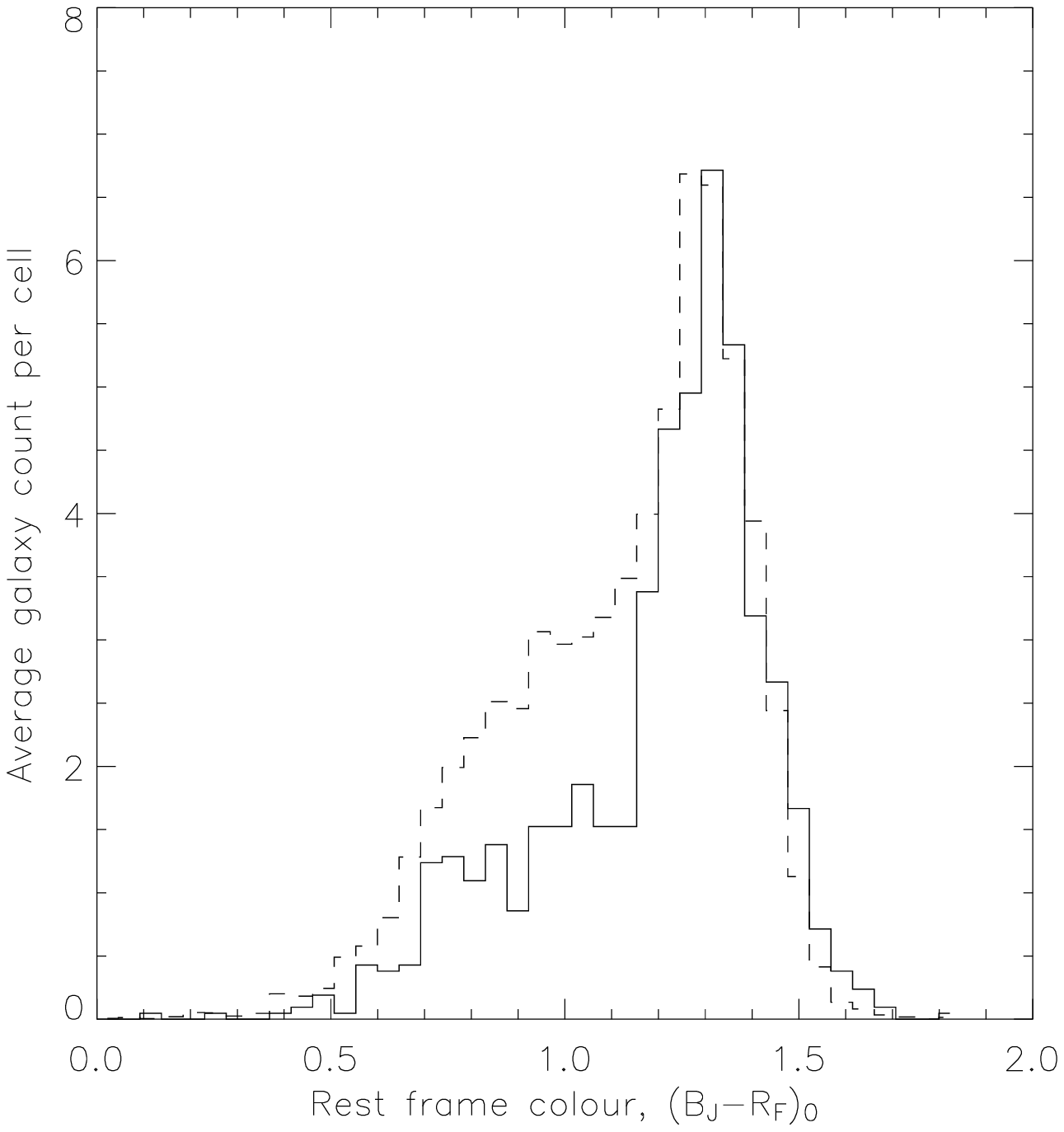}
\end{minipage}
\hspace{0.6cm}
\begin{minipage}{8.0cm}
\includegraphics[scale=0.6]{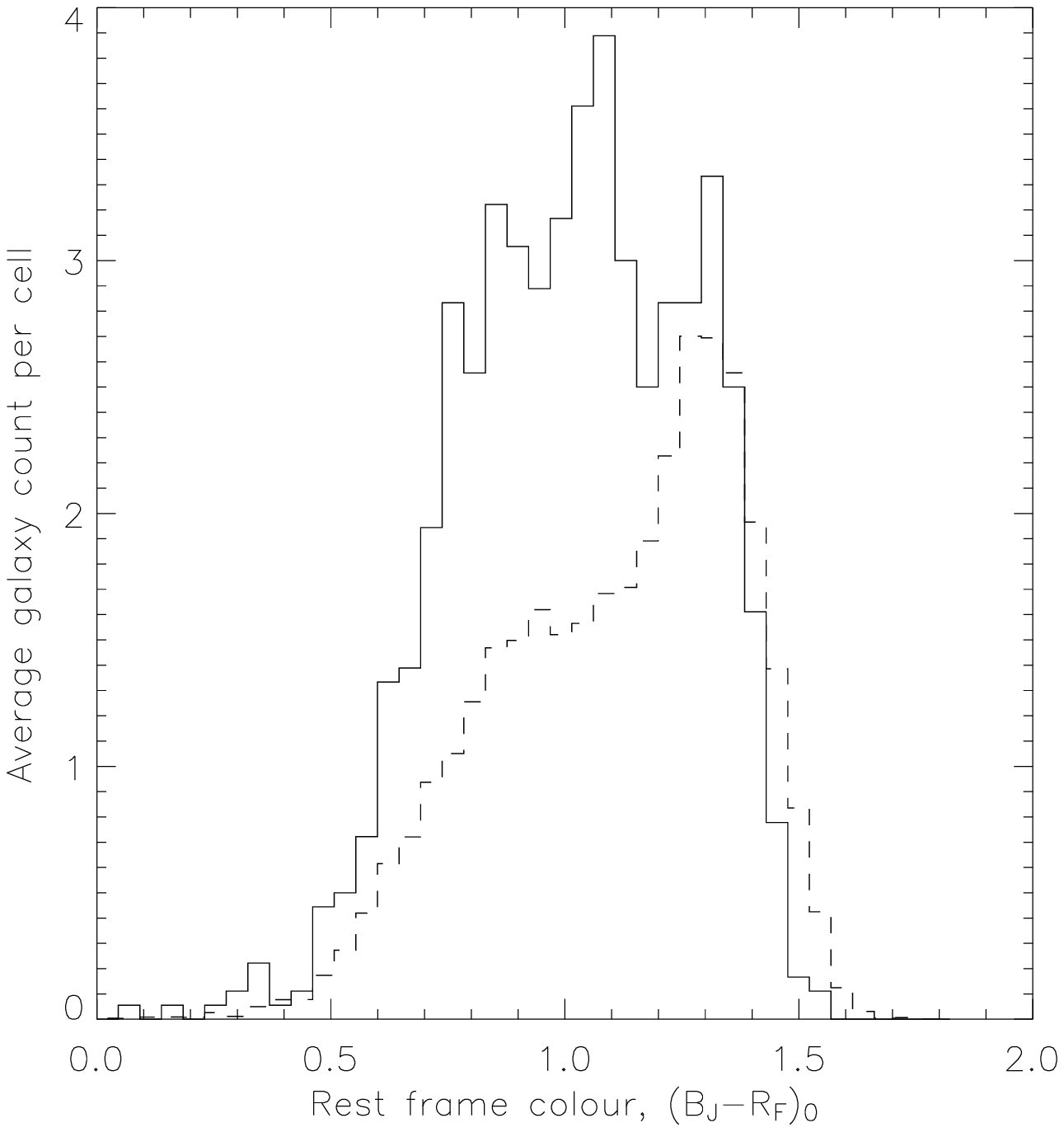}
\end{minipage}
\caption{\small The colour distribution of cells identified as causing
  the stochasticity signal ($L=20\Mpc$, filled line). The left (right) plot shows all those cells with
  an excess of red (blue) galaxies. The comparison plot (dashed line)
  is calculated from those cells with similar $\delta_\Ej$ to the
  outlier cells, in order to account for nonlinearities.
  }
\label{fig_coldist}
\end{minipage}
\end{figure*}

Before the detection of stochastic bias is accepted, and
we proceed to confront the result with theoretical models,
a degree of skepticism is in order. We have seen 
that some regions of space have a number ratio
of early and late type galaxies that differs from
the typical value by too much to be consistent with
Poisson scatter. Such an outcome seems potentially
vulnerable to systematics in the analysis as any source
of error in classifying galaxies could introduce an
extra scatter, spuriously generating the impression
of stochasticity. However, it is not clear which way
this effect would go. Suppose the survey finds galaxies
with perfect efficiency, but then assigns them a random
class. 
Any true initial stochasticity is erased by the classification
`errors' and we measure $r_\LN=1$.
In order to generate apparent stochasticity where
none is present we would need something more subtle.
Possibilities could include a perfect efficiency in detecting early type
galaxies, but a fluctuating efficiency in finding late types; a
spatially varying  boundary between early and late types; or large
variations in the survey selection function on scales smaller than the
cell length. To assess the possible contribution of this latter effect
to our measured stochasticity, we applied small scale incompleteness
masks to semi-analytic datasets (see Section \ref{sec_dur}). The large
scale stochasticity of these models was affected by less than 1 sigma. 

Whether or not a spurious generation of stochasticity
seems plausible, it is worth looking more closely
at the data to see how the signal arises.
In order to do this, we focus on the outliers
from the relation $\ln(1+\delta_\Lj) \propto \ln(1+\delta_\Ej)$, but a careful
definition of an outlier is required. We want to ask how much the
numbers $(N_\Ej,N_\Lj)$ differ from their expectation
values when clustering is included, but the latter are unknown. Therefore, we take
the best-fitting power law model with $r_\LN=1$ and
integrate over the distribution of densities to calculate
the probability for obtaining this
outcome, $P(N_\Ej,N_\Lj)$, accounting for Poisson noise. The most outlying
points are those with the lowest values of $P$, and
we remove these in succession until the remaining cells are
consistent with an $r_\LN=1$ model. The numbers of
outliers in this sense are listed in Table \ref{tab2}, and
Fig. \ref{fig_colsd} shows their positions on density-density plots.

Having identified the cells that provide the evidence for
stochasticity, we can examine their properties in more detail.
Fig. \ref{fig_outliers} shows the
spatial distribution of the outlying cells within the 2dFGRS for
a range of cell sizes,
from which it can be seen that they are often associated with
overdense regions. This should not be taken as indicating that
stochasticity is confined to such regions: given that the
degree of stochasticity is small, the cells that contain the
most galaxies will provide the best signal-to-noise for the
effect. The colour distribution of galaxies in the
outlying cells is shown in Fig. \ref{fig_coldist}, compared to the
distribution of `normal' cells.
To allow for nonlinearity in the density-density relation we consider
for the comparison distribution only those cells with similar values
of $\delta_\Ej$. 
The distributions cover sensible ranges of colours,
and the peaks corresponding to early and
late types appear to be in the correct places. What causes these
cells to be outliers is that the ratio of the
two populations differs greatly from what is typical, and
it is hard to see how this result can be in error.
The completeness values in these cells are typically
0.8, and yet we see variations in the early:late ratio by
more than a factor 2. Moreover, similar variations
are seen whether we classify using colour or spectral type.
We therefore conclude that these variations are a real property
of the galaxy distribution.

\subsection{Consistency checks} \label{sec_MC}
We repeat the analysis for cells with galaxies
classified randomly,  recovering a best--fitting bivariate lognormal
model with $r_\LN = 1$ exactly.
By fitting the bivariate lognormal model to Monte Carlo simulated
power law mocks (Section \ref{sec_modelcomp}), we can check for any
bias inherent in our fitting procedures. The best fit models
have mean $r_{\LN} \gsim 0.998$, not significantly different from the
$r_{\LN}= 1$ of power law deterministic bias.

\subsection{Direct variance estimates}\label{efst90}

It is possible to determine the variance $\sigma^2(L)$
directly without assuming the lognormal model. Optimal power spectrum
estimates perhaps provide the most accurate determinations of variance
\citep{2004ApJ...606..702T,2003MNRAS.346..994P}, however for our
purposes it suffices to use a simpler method presented by
\citet{1990MNRAS.247P..10E}. Their estimator
calculates $\Delta N = N-\langle N \rangle$
for each cell and subtracts the Poisson variance from
$(\Delta N)^2$ to form an estimate of $\sigma^2$ for
each cell. This is then averaged over all cells.
The estimator only applies in the case of a uniform survey, where
$\langle N \rangle$ is the same for all cells. For the general case of
an incomplete survey, Efstathiou et al. derive a slightly different
estimator, assuming a Gaussian density
field. In fact, this is a
poor assumption even on the largest scales considered here. We use
Monte Carlo realisations of lognormal fields to show that their
estimator for $\sigma$ is biased low by around 1-2\%, and has an
uncertainty often several times larger than that expected for the Gaussian
model. Table \ref{tab6} gives the direct variance estimates for our
data, with errors from both \citet{1990MNRAS.247P..10E} and Monte Carlo
simulations. 

\begin{table*}
\centering
\parbox{15.75cm}{\caption{\label{tab6} \small Different variance estimates and
    errors for early and late datasets defined by colour. (a) From
    our bivariate lognormal model fit with errors derived from a
    multidimensional Gaussian fit to the likelihood surface; (b)
    Efstathiou et al. (1990) direct variance estimator and errors; (c)
    direct variance estimator after using Monte Carlo simulations of
    lognormal fields to correct for bias due to non-Gaussianity, with
    rms errors from the simulations.}}
\vspace{0.2cm}
\begin{tabular}[htb]{ccccccccccccc} \hline \hline
  cell size & 
  ${\sigma_\Ej}^{\rm a}$ & ${\Delta(\sigma_\Ej)}^{\rm a}$ &
  ${\sigma_\Ej}^{\rm b}$ & ${\Delta(\sigma_\Ej)}^{\rm b}$ &
  ${\sigma_\Ej}^{\rm c}$& ${\Delta(\sigma_\Ej)}^{\rm c}$ &
  ${\sigma_\Lj}^{\rm a}$& ${\Delta(\sigma_\Lj)}^{\rm a}$ &
  ${\sigma_\Lj}^{\rm b}$ & ${\Delta(\sigma_\Lj)}^{\rm b}$ &
  ${\sigma_\Lj}^{\rm c}$& ${\Delta(\sigma_\Lj)}^{\rm c}$ \\  \hline

10 & 3.014 & 0.074 & 1.806 & 0.016 & 1.814 & 0.142 & 1.799 & 0.037 & 1.321 & 0.013 & 1.323 & 0.059 \\
15 & 1.987 & 0.060 & 1.462 & 0.021 & 1.472 & 0.129 & 1.295 & 0.033 & 1.054 & 0.016 & 1.056 & 0.056 \\
20 & 1.542 & 0.057 & 1.242 & 0.028 & 1.253 & 0.126 & 1.024 & 0.035 & 0.879 & 0.021 & 0.882 & 0.055 \\
25 & 1.214 & 0.048 & 1.008 & 0.034 & 1.019 & 0.107 & 0.839 & 0.032 & 0.746 & 0.026 & 0.750 & 0.054 \\
30 & 0.922 & 0.045 & 0.877 & 0.042 & 0.890 & 0.102 & 0.677 & 0.033 & 0.659 & 0.032 & 0.663 & 0.061 \\
35 & 0.807 & 0.064 & 0.728 & 0.041 & 0.736 & 0.087 & 0.584 & 0.042 & 0.547 & 0.032 & 0.551 & 0.050 \\
40 & 0.764 & 0.067 & 0.716 & 0.049 & 0.727 & 0.095 & 0.596 & 0.049 & 0.557 & 0.038 & 0.563 & 0.061 \\
45 & 0.657 & 0.054 & 0.547 & 0.047 & 0.557 & 0.071 & 0.546 & 0.044 & 0.467 & 0.041 & 0.473 & 0.058 \\

\end{tabular}
\end{table*}

Even accounting for the bias, these direct variance estimates remain
generally 10--20\% lower than those estimated by fitting a Poisson sampled
lognormal curve. For early type galaxies in $L=10\Mpc$ cells the
discrepancy is nearly 40\%. Imposing different weighting schemes
during fitting can lower the lognormal variance to meet the direct variance results
(Conway et al. 2004), but
these have no significant effect on our measurement of stochasticity. 

This failure of the lognormal model to recover the true variance of the data
may be due to the assumption of the {\it Poisson
Clustering Hypothesis\/} which we know to be incorrect in detail. On
the smallest scales our cells are largely shot noise dominated, and it
is on these scales that the discrepancy is greatest (see also Section
\ref{sec_1dfailure}).  
It remains important to emphasise that the variance estimates given
throughout this paper are model dependent, and not to be
taken as the true variance of the galaxies in the survey, which can be
estimated more accurately through model independent methods. 

An unfortunate side effect of this difficulty in obtaining accurate
estimates for variance, is that the average biasing statistics of
Section \ref{sec_frame} are dependent on $\sigma$ (see also
Section \ref{sec_scale}). Tests show
this to have little effect on stochasticity ($\sigma_b/\hat{b}$),
but on scales $\le 20\Mpc$ nonlinearity is
overestimated. By replacing our measured variance with results
obtained from bias corrected direct estimation we find nonlinearity to
decrease to around 2\% for $L=10\Mpc$, decreasing with scale gradually
to match our measured values by $L=30\Mpc$. Stochasticity is decreased
by about 2 sigma at $L=10\Mpc$ to around 0.39, but the effect is
insignificant on all other scales. 
These results may be compared with the measurements of variance and
deterministic bias in the 2dFGRS using flux-limited samples over a
slightly larger volume (Conway et al. 2004).

This is a suitable point to discuss a subtlety of 
cell counts that we have neglected so far. 
Variation in the survey mask is represented by $\langle N \rangle$
varying between cells. We have treated this as a simple variation in
sampling efficiency that is uniform over the cell. However,
this cannot be precisely correct: where sampling of a cell
is low because it encounters one of the larger drills in the
input catalogue, it would be more correct to assume a
completely sampled cell of smaller volume. We have explored
this alternative extreme by assuming that $\sigma \propto \langle N \rangle^{-0.3}$,
as expected for a $\xi(r)\propto r^{-1.8}$ spectrum. 
Since $\langle N \rangle \propto$ completeness, and the typical cell
completeness is about 0.8, the measured values of $\sigma$ are
increased by about 7~per~cent, approximately a 1-$\sigma$ shift.
This has no effect on our detection of stochasticity, and because the
`lost volume' assumption will not apply in all cases, we neglect the issue. Note that estimates
of cell variances derived from integration over correlation functions
or power spectra would be completely independent of this issue.

\section{Comparison with simulations}\label{sec_sims}

In order to interpret our measurements of stochastic bias, we
need to make a comparison with theory. In practice,
this means considering the results of numerical simulations that
are sufficiently detailed to predict the spatial distributions of the
different classes of galaxy.
There are currently two main methods of simulating the large scale
structure of the visible universe: semianalytic or hydrodynamic. 
We consider each in turn.

\subsection{Previous work}

\citet{2001MNRAS.320..289S} used semianalytic models to
measure the relative bias between early and late type galaxies (as
defined by bulge to total luminosity) and red and blue galaxies on
scales of $r=8\mpcoh$. They set a limiting
magnitude of $M_B -5\log\,h \leq - 18.4$, and split galaxies by colour at
$B-R = 0.8$, making their samples reasonably comparable to our data
for cells of $L=20\Mpc$. 
Our value of $b_{\rm var} = 0.66$ (Table \ref{tab3}) falls in between their values of 
 $0.77$ for late/early types and $0.55$ for blue/red galaxies. 
They find $r_{\rm lin} = 0.87$ for both
subgroups, slightly lower than our values for both colour and
spectral type. Unfortunately the results are not split into
stochasticity and nonlinearity, making it difficult to make further
comparisons. 
It is  however interesting that we find a lower amplitude
of relative bias between the two colour groups than is seen in these models.

The hydrodynamic simulations of
\citet{2001ApJ...558..520Y} classify galaxies by their
formation redshift, and are smoothed with top hat spheres of radius $8\mpcoh$. 
By using this classification scheme,
hydrodynamic models approximate early type galaxies as those that form at high
redshifts via initial starbursts, whereas late type galaxies have 
a lower formation redshift and undergo slower star formation.
They find that old galaxies are positively biased with respect
to matter with a linear correlation coefficient of less than 1,
whereas young galaxies are slightly antibiased with a correlation
coefficient closer to 1. They
measure the relative bias between galaxy types by $b^{\rm rel}_\xi \equiv
(\xi_{\rm young}/\xi_{\rm old})^{1/2}$ where $\xi_{\rm young}$ ($\xi_{\rm old}$) is
the two-point correlation function of the young (old) galaxies. This
is equivalent to $b_{\rm lin}$. They obtain values of
between 0.5 and 0.66 for scales of $1\mpcoh < r < 20\mpcoh$,
lower than our equivalent values for the linear biasing model with
$L\le 25\Mpc$ cells (Table \ref{tab2}). Once again results for stochasticity
and nonlinearity are not quoted for the relative bias.

\subsection{Preliminary mock comparison}\label{sec_dur}

None of this past work really allows a direct comparison
with our results, so we generated two new
theoretical `datasets' from the results of large
semianalytic calculations carried out using the `Cosmology machine'
supercomputer at Durham. The background model is that deduced
from the simplest WMAP+2dFGRS analysis of \citet{2003ApJS..148..175S}:
flat, $\Omega_m=0.27$, $\Omega_b=0.045$, $h=0.72$, $n=0.97$,
$\sigma_8=0.8$, applying the semianalytic apparatus of
\citet{2000MNRAS.319..168C} to a simulation with $N=500^3$ particles
in a box of side $250\mpcoh$. As shown by e.g. \citet{2003ApJ...599...38B},
a problem faced by such modelling is a tendency to over-produce
massive galaxies, as a result of excessive cooling arising
from the higher baryon density now required by CMB+LSS. This
problem is particularly severe for disk (late type) galaxies.
The first mock adopted the `superwind' approach of
\citet{2003ApJ...599...38B} in an attempt to alleviate this problem,
but the cure is not total. The second mock attempted to reduce cooling
by retaining the low baryon density of
\cite{2000MNRAS.319..168C}. Although this
conflicts with CMB data, it provides a useful means of
comparison. For this application, we took an empirical approach in which a
monotonic shift in luminosity was applied to force the models to
have the observed luminosity function as in \citet{2002MNRAS.333..133M}.
The model colour distribution was bimodal to a realistic degree,
so this shift was applied separately to generate model distributions
of early and late type galaxies in which the global luminosity
functions were correct. 
The resultant mock cell counts were analysed identically to the real
data.

In some respects, these simulations match the real data
very well. For the low-baryon model, the amplitude of the cell
variances for early-type galaxies agree to within 3\% on small scales
and 10\% on large scales. The superwind model variances agree to within 10\% on
all scales. The relative bias of the low-baryon model agrees to
within 10 and 15\%  with observation, and the superwind model to
within 10 and 20\%. 
Significant stochasticity and nonlinearity is also required, 
which can be measured accurately as a function of scale
because we are able to use more mock cells than are available
in the real data. The mocks are affected in a similar manner to the
data by the discrepancy between direct estimates of variance and those
from lognormal model fits. On small scales this significantly
increases our estimated nonlinearity; as the effect is
equivalent between mocks and data, however, a direct comparison between them
remains instructive. Fig. \ref{fig_nonstoc} shows the resulting
stochasticity and nonlinearity as a function of scale, compared to
that of the 2dFGRS data.
The impression is that the mock results
show a greater nonlinearity than the real data on small scales, while
stochasticity is well matched within the errors.

Given the known imperfect nature of the semianalytic simulations
(e.g. the failure to match luminosity functions exactly), the
correct attitude is probably to be encouraged by the degree
of agreement with the data. It is certainly plausible that
the existing calculations contain all the relevant physical
contributions to bias, but perhaps not yet in quite the
right proportions.
As usual with such numerical comparisons, this raises the
question of whether the issue of stochasticity can be understood
in a more direct fashion. In the end, the effects we are seeing must
be reducible to the way in which the early:late ratio varies
between and within virialized systems of different mass, so that in
effect we are dealing with a more general version of the
morphological segregation that is familiar from the study of rich clusters
\citep*{2000ApJ...528....1N}.
We intend to pursue this in more detail elsewhere, using the
catalogue of galaxy groups derived from the 2dFGRS by \citet{2004MNRAS.348..866E}.

\section{Summary and Conclusions}

We have presented fits of three relative biasing schemes to
joint counts-in-cells distributions of 2dFGRS galaxies, separated by
both colour and spectral type
$\eta$. Each scheme is convolved with a Poisson distribution to account
for statistical `shot noise'. Our first two models present two
alternative types of deterministic biasing: linear and power law
bias. Linear bias is an important concept in cosmology and many
results are linked to it, but it is not physically plausible as it 
allows negative densities. Power law bias presents a simple cure for
this problem, but still has little physical motivation. 
With the advent of large semianalytic and hydrodynamic simulations,
interest has grown in `stochastic' bias models. Bias could be
determined by parameters other than the local overdensity of the dark
matter, and considerable scatter could occur in the
relation. Galaxy distributions have previously been measured to be well approximated as
lognormal, therefore a bivariate lognormal distribution seems a natural model
for relative bias between galaxy types. This model
incorporates stochasticity and nonlinearity in a well defined
manner, which is mathematically simple and consistent with observation.

To account for the discrete nature of galaxies, the {\it Poisson
sampling hypothesis} is assumed, and all models are convolved with a
Poisson distribution. On small scales where our cell
counts become shot noise dominated, we find this hypothesis to fail
causing overestimates of variance compared to direct estimation
methods. The main symptom of the discrepancy is a number of completely
empty cells that exceeds the Poisson sampled lognormal prediction.
This is found not to affect our results for stochasticity, and the
same effect is seen in the simulations, 
but it emphasises the need for a greater understanding of Poisson
statistics in relation to galaxy clustering.

We have detected a significant deviation from  $r_\LN=1$ in the 2dFGRS and
confirmed this detection of stochasticity through likelihood ratio tests,
Kolmogorov-Smirnoff probability testing, and Monte Carlo
simulations. We have measured stochasticity at a level of
$\sigma_b/\hat{b}=0.44 \pm 0.02$ or $r_\LN=0.958 \pm 0.004$
on the smallest scales ($10\Mpc$), declining with increasing cell size. The
nonlinearity of the biasing relation is less than $5\%$ on all scales.
The small measured values of
stochasticity and nonlinearity support the use of galaxy redshift
surveys for studies of the large scale distribution of matter in the
universe, and the measurement of cosmological parameters. However, as 
precision in cosmology increases and new techniques are developed, the
effects of stochastic bias on parameter estimation should be
understood. For example, studies of cosmology through weak
gravitational lensing requires knowledge of nonlinear and stochastic 
bias \citep{astro-ph/0403698}. Our results 
for $r_{\rm lin}$ on 10~Mpc scales are consistent within the (large)
errors with galaxy-mass correlations measured by weak lensing surveys
\citep{2002ApJ...577..604H} on the largest scales probed.

A comparison with semianalytic simulations shows a similar
variation of nonlinearity and stochasticity with scale. The
amplitude of stochasticity appears a little lower than in the true
data, particularly on large scales, and the nonlinearity is slightly greater
on small scales. Nevertheless, given the known imperfections of the
current generation of semianalytic calculations, the general
agreement is certainly encouraging.
We hope that this work will stimulate the investigation of
more detailed biasing models.
Through the linking of new simulations to observations, a more
thorough understanding of the processes of galaxy formation and
evolution should be within our reach.

\section*{acknowledgements}
The 2dF Galaxy Redshift Survey was made possible through the dedicated 
efforts of the staff of the Anglo-Australian Observatory, both in creating 
the 2dF instrument and in supporting it on the telescope.
JAP and OL thank PPARC for their Senior Research Fellowships.
Many thanks to Paul Hewett, Yehuda Hoffman, Mike Irwin and Dan Mortlock for helpful
discussions on galaxy evolution and statistics.

\bibliographystyle{mn2e}


\begin{appendix}
\section{Non-linear and stochastic bias statistics for the Bivariate
  Lognormal Distribution}\label{app1}

The general biasing relation between the overdensities
$\delta_1$ and $\delta_2$ of two subgroups of galaxies, or two types
of matter, is fully described by equation (\ref{eq_b})
\be\label{A1}
b(\delta_1)\delta_1 \equiv \langle\delta_2|\delta_1\rangle = \int f(\delta_2|
\delta_1)\delta_2 d\delta_2.
\ee
The conditional probability distribution for the bivariate lognormal
model is given by equation (\ref{eq_bivLNcond})
\be\label{A2}
f(g_2| g_1) =
\frac{\omega_1}{(2\pi|V|)^{1/2}}\exp\left[-\frac{(\tilde{g_2}-r_{\LN}\,\tilde{g_1})^2}{2(1-r_{\LN}^2)}\right]
\ee
where  $g_i = \ln(1+\delta_i) - \langle \ln(1+\delta_i)\rangle$, $\langle
\ln(1+\delta_i)\rangle = -\omega_i^2/2$ and $\tilde{g_i} =
g_i/\omega_i$. $\tilde{g_2}|\tilde{g_1}$ follows a univariate Gaussian distribution with mean
$r_{\LN}\,\tilde{g_1}$ and
variance $1-r_{\LN}^2$. The covariance matrix $V$ and correlation coefficient 
$r_{\LN}$ are both defined in log space, and are given explicitly in 
equations (\ref{eq_rln}) and (\ref{eq_V}). The variance of the
distribution in linear space $\sigma_i^2$ is related to the variance of the
Gaussian field by
\be 
\sigma_i^2 \equiv \langle \delta_i^2 \rangle = \exp[\omega_i^2]-1.
\ee

On substituting equation (\ref{A2}) into (\ref{A1}) and integrating we find
\be
b(\delta_1)\delta_1 = \exp\left[ \omega_2
\tilde{g_1} - \frac{(\omega_2 r_{\LN})^2}{2}\right] -1.
\ee
From this basic parameter we can calculate the mean biasing and its
nonlinearity [equations (\ref{eq_hatb}) and (\ref{eq_tildeb})]
\be
\hat{b} \equiv \frac{\langle b(\delta_1) \delta_1^2
\rangle }{\sigma_1^2} = \frac{\exp[r_{\LN} \omega_2 \omega_1]-1}{\exp[\omega_1^2]-1}
\ee
\be
\tilde{b}^2 \equiv \frac{\langle b^2(\delta_1) \delta_1^2
\rangle }{\sigma_1^2} = \frac{\exp[r_{\LN}^2 \omega_2^2]-1}{\exp[\omega_1^2]-1}
\ee
We also know the ratio of variances, equation (\ref{eq_bvar})
\be
b_{\rm var}^2 \equiv \frac{\sigma_2^2} {\sigma_1^2} =
\frac{\exp[\omega_2^2]-1}{\exp[\omega_1^2]-1}.
\ee

Although it is possible to derive the scatter for this model from
equation (\ref{eq_sigb}), it can be shown that (DL99)
\be
b_{\rm var}^2 = \tilde{b}^2 + \sigma_b^2.
\ee
Using this fact we obtain
\be
\sigma_b^2 = \frac{\exp[\omega_2^2] - \exp[r_{\LN}^2 \omega_2^2]}{\exp[\omega_1^2]-1}.
\ee

Whilst the bivariate lognormal model contains constant scatter dependent only on
$r_{\LN}$ in the log frame, transformation to the linear frame causes
the the scatter to become dependent on the
widths of the univariate distributions and vary with $\delta_1$.

\end{appendix}

\end{document}